\documentclass[11pt,english]{article}
\pdfoutput=1 
\usepackage{jheppub} 
\usepackage[T1]{fontenc}
\usepackage[latin9]{inputenc}
\usepackage{color}
\usepackage{babel}
\usepackage{amssymb}
\usepackage{amsmath}
\usepackage{enumitem}
\usepackage{graphicx}
\usepackage{mathtools}
\usepackage{ragged2e}
\usepackage{epsfig}
\usepackage{bm}
\usepackage{xcolor}
\usepackage{times}
\usepackage[export]{adjustbox}
\makeatletter
 \definecolor{BLACK}{gray}{0}
 \definecolor{WHITE}{gray}{1}
 \definecolor{RED}{rgb}{1,0,0}
 \definecolor{GREEN}{rgb}{0,0.5,0}
 \definecolor{BLUE}{rgb}{0,0,1}
 \definecolor{CYAN}{cmyk}{1,0,0,0}
 \definecolor{MAGENTA}{cmyk}{0,1,0,0}
 \definecolor{YELLOW}{cmyk}{0,0,1,0}
\makeatother

\title{Out-of-horizon correlations following a quench in a relativistic quantum field theory}
\author[a,b]{I. Kukuljan}
\author[c]{S. Sotiriadis} 
\author[d]{G. Tak\'acs}
\affiliation[a]{Max-Planck-Institute of Quantum Optics, Hans-Kopfermann-Str. 1, DE-85748 Garching, Germany}
\affiliation[b]{Munich Center for Quantum Science and Technology, Schellingstr. 4, DE-80799 M{\"u}nchen, Germany}
\affiliation[c]{University of Ljubljana, Faculty of Mathematics and Physics, Jadranska 19, SI-1000 Ljubljana, Slovenia}
\affiliation[d]{BME ``Momentum'' Statistical Field Theory Research Group \& BME Department of Theoretical Physics, H-1117 Budapest, Budafoki \'ut 8, Hungary}

\emailAdd{ivan.kukuljan@mpq.mpg.de}
\emailAdd{spyridon.sotiriadis@fmf.uni-lj.si}
\emailAdd{takacsg@eik.bme.hu}

\abstract{
One of the manifestations of relativistic invariance in non-equilibrium quantum field theory is the ``horizon effect'' a.k.a. light-cone spreading of correlations: starting from an initially short-range correlated state, measurements of two observers at distant space-time points are expected to remain independent until their past light-cones overlap. Surprisingly, we find that in the presence of topological excitations correlations can develop outside of horizon and indeed even between infinitely distant points. We demonstrate this effect for a wide class of global quantum quenches to the sine-Gordon model. We point out that besides the   maximum velocity bound implied by relativistic invariance, clustering of initial correlations is required to establish the ``horizon effect''. {We show that quenches in the sine-Gordon model have an interesting property: despite the fact that the initial states have exponentially decaying correlations and cluster in terms of the bosonic fields, they violate the clustering condition for the soliton fields, which is argued to be related to the non-trivial field topology. The nonlinear dynamics governed by the solitons makes the clustering violation manifest also in correlations of the local bosonic fields after the quench.}
}

\begin{document}
\maketitle

\section{\textbf{Horizon effect in quantum field theory}} 
Consider the following thought experiment: A system described by a quantum field theory (QFT) is prepared in some initial state and let to evolve under unitary relativistic dynamics. This can be realised by a paradigmatic protocol known as a ``quantum quench'' \cite{CC}, in which a closed quantum system initially prepared in an equilibrium state undergoes a sudden change of its Hamiltonian at time $t=0$. Let us suppose that the initial state is characterised by short-range correlations of local quantum fields with a small correlation length $\xi$; such states are rather common and include ground and thermal states of massive QFTs. As a result, measurements made by two observers separated by a distance $r$ will be independent until time $t\sim r/2c$, where $c$ is the finite maximum speed at which information propagates \cite{LC_Igloi,CC}. This is known as the "horizon effect" and can be justified by a semiclassical interpretation: correlations propagate by pairs of entangled quasiparticles emitted from initially correlated nearby points (at a distance $\lesssim\xi$) that travel to opposite directions with velocities limited by $c$. The horizon effect can be demonstrated in Conformal Field Theory for a class of initial states \cite{CC,CC2007,Cardy2016}, and can be proved in the case of non-interacting relativistic dynamics for all initial states that satisfy the cluster decomposition property as shown in App.~\ref{sec:Horizon_freeQFT}. Several works have verified the presence of horizon both analytically and numerically also in lattice systems with local Hamiltonians \cite{LCTheor1,LCTheor2,CalabreseTransverseIsing2011,CalabreseTransverseIsingI2012,CalabreseTransverseIsingII2012,EsslerLocalQuenches2012,EsslerQuench2012,LCTheor3,LCTheor4,Krutitsky2014,BertiniLightcone2016}, whose dynamics,
even though not relativistically invariant, is still constrained by
a maximum velocity of information propagation \cite{LRbound}. The effect has been observed in disordered systems \cite{CalabreseChiara2006,Burrell2007,Igloi2012,BardarsonMoore2012,Altman2014,Altman2015,Zhao2016} and systems with long-range interaction \cite{Hauke2013,Schachenmayer2013,Richerme2014,SotiriadisLogRange2015,BuyskikhEssler2016}, and has also been found for the time evolution of the entanglement entropy \cite{CalabreseCardy2005,CalabreseChiara2006,LCTheor1,FagottiCalabrese2008,Eisler2008,Kim2013,Nezhadhaghighi2014,ColluraCalabrese2014,CalabreseQuenchExcitedStates2014,CalabreseQuenchExcitedStatesII2014,FagottiPrethermalisation2015}.  
The horizon effect has also been observed in cold-atom experiments \cite{LCexp1,LCexp1a,LCexp2}. Based on these results, the horizon effect has been generally accepted as a universal feature of non-equilibrium quantum field theory and non-equilibrium quantum statistical physics in general.

It is known that non-trivial interactions can have significant effects on the speed of propagation \cite{LCTheor4} or even fully suppress the spreading of correlations \cite{Isingconfinement}. However, their interplay with the horizon effect remains incompletely understood. Here we address this question in the context of the sine-Gordon (SG) model described by the Hamiltonian (using units in which $c=1$)
\begin{equation}
H_{SG}=\int\left(\frac{1}{2}{\Pi}{}^{2}+\frac{1}{2}(\partial_{x}\Phi)^{2}- \frac{\mu^2}{\beta^2}\cos\beta\Phi\right)\mathrm{d}x\label{eq:SG}
\end{equation}
where $[\Phi(x),\Pi(x')]=\mathrm{i}\delta(x-x')$, $\mu$ is the mass parameter and $\beta$ the interaction parameter of the theory. The SG model is a prototypical example of a 1+1 dimensional relativistic QFT with rich physics governed by topological excitations called  solitons and antisolitons. They appear due to the periodic cosine potential compactifying the field to a circle topology  $\Phi\sim\Phi+2\pi/\beta$. Moreover, when the coupling parameter $\beta$ is in the attractive regime $\beta^2<4\pi$, they form neutral bound states called breathers. The SG model has a non-trivial phase diagram with a transition of the Berezinskii-Kosterlitz-Thouless type at $\beta^2=8\pi$. It is an integrable model \cite{Zamo77,Zamo-Zamo} and describes the dynamics of numerous condensed matter systems \cite{Giamarchi}. The SG model has recently been realised in an ultra-cold atom experiment which enables the study of its correlation functions and non-equilibrium dynamics \cite{exp-sG, phase_locking}. Its dynamics has attracted considerable recent interest \cite{semiclassicaldynamics,Iucci2010,Cubero-Schuricht,Horvath-Takacs, universalrephasing, sGFF_evolution, semicl1, semicl2,twatcsa,Rylands2019}.

We prepare the system in an initial state $|\Omega\rangle$ with finite correlation length $\xi$ and clustering property for local boson fields. The canonical case is the ground state of the massive Klein-Gordon (KG) Hamiltonian 
\begin{equation}
H_{KG}=\int\left(\frac{1}{2}\Pi{}^{2}+\frac{1}{2}(\partial_{x}\Phi)^{2}+\frac{1}{2}m_{0}^{2}\Phi^{2}\right)\mathrm{d}x
\end{equation}
for which $\xi=1/m_0$, while another case is the ground state of the SG model with parameters $\mu_0$, $\beta_0$.
At time $t=0$ we switch on SG dynamics with parameters $\mu$, $\beta$ and study the time evolution of connected correlations 
\begin{equation}
C_{\mathcal{O}}(x,y;t)\coloneqq\langle\Omega|\mathrm{e}^{+\mathrm{i}H_{SG}t}\mathcal{O}(x)\mathcal{O}(y)\mathrm{e}^{-\mathrm{i}H_{SG}t}|\Omega\rangle_{\text{conn}}
\end{equation}
of local observables $\mathcal{O}$. 
We focus on three physically relevant observables: the field $\Phi(x)$ and its derivatives $\partial_{x}{\Phi}$ and $\partial_{t}{\Phi}=\Pi$ corresponding to the soliton density and current respectively. Although $\Phi(x)$ is an angular variable, its correlations are experimentally observable \cite{exp-sG}.

Using numerical simulations based on the Truncated Conformal Space Approach (TCSA) \cite{Yurov-Zamo,TCSA-review}, we observe violation of the horizon effect and long-range correlations that extend through the whole system and exhibit oscillatory time dependence. The violation turns out to be a general feature of quenches to the sine-Gordon model, it is robust under changes in the parameters of the dynamics or the initial state, and is strongly sensitive to the boundary conditions. Exploiting the mapping between sine-Gordon and massive Thirring model~\cite{Coleman} based on Bosonisation~\cite{Delft,GNT-bosonization,Langmann2015}, we analytically verify this numerical observation by exactly solving the full quench dynamics at the point $\beta=\sqrt{4\pi}$ (free fermion point). Our calculation shows that the horizon violation can be attributed to the topological nature of the soliton fields~\cite{Mandelstam}.

The paper is organised as follows. 
In sec.~\ref{sec:surprise}, we present a numerical investigation of horizon dynamics in the SG model and the surprising finding of the horizon violation
In sec.~\ref{sec:demonstration} we outline the analytical solution for the horizon violation at the free fermion point. Based on this derivation, in sec.~\ref{sec:explanation} we propose a physical explanation of the phenomenon. Further analysis of the characteristic properties of the horizon violation is presented in sec.~\ref{sec:FurtherProperties}. In sec.~\ref{sec:experiment}, we propose a protocol to measure the horizon violation in ultra-cold atom experiments.
Lastly, in sec.~\ref{sec:conclusions} we summarise our conclusions, comment on the exceptionality of our findings within the general framework of QFT and discuss outlook for further study. Further details on methods used in this work are presented in the appendices. App.~\ref{sec:Horizon_freeQFT} contains a proof of the horizon effect in the case of free dynamics and initial states that are local in terms of the freely evolving fields. App.~\ref{app:TCSA} provides technical details of the numerical simulations. Lastly, app.~\ref{app:BF} provides a more detailed presentation of each of the steps of the analytical solution.

\section{{A surprise: out-of-horizon spreading of correlations}}\label{sec:surprise}

\begin{figure*}[!ht]
\centering{\includegraphics[clip,width=\textwidth]{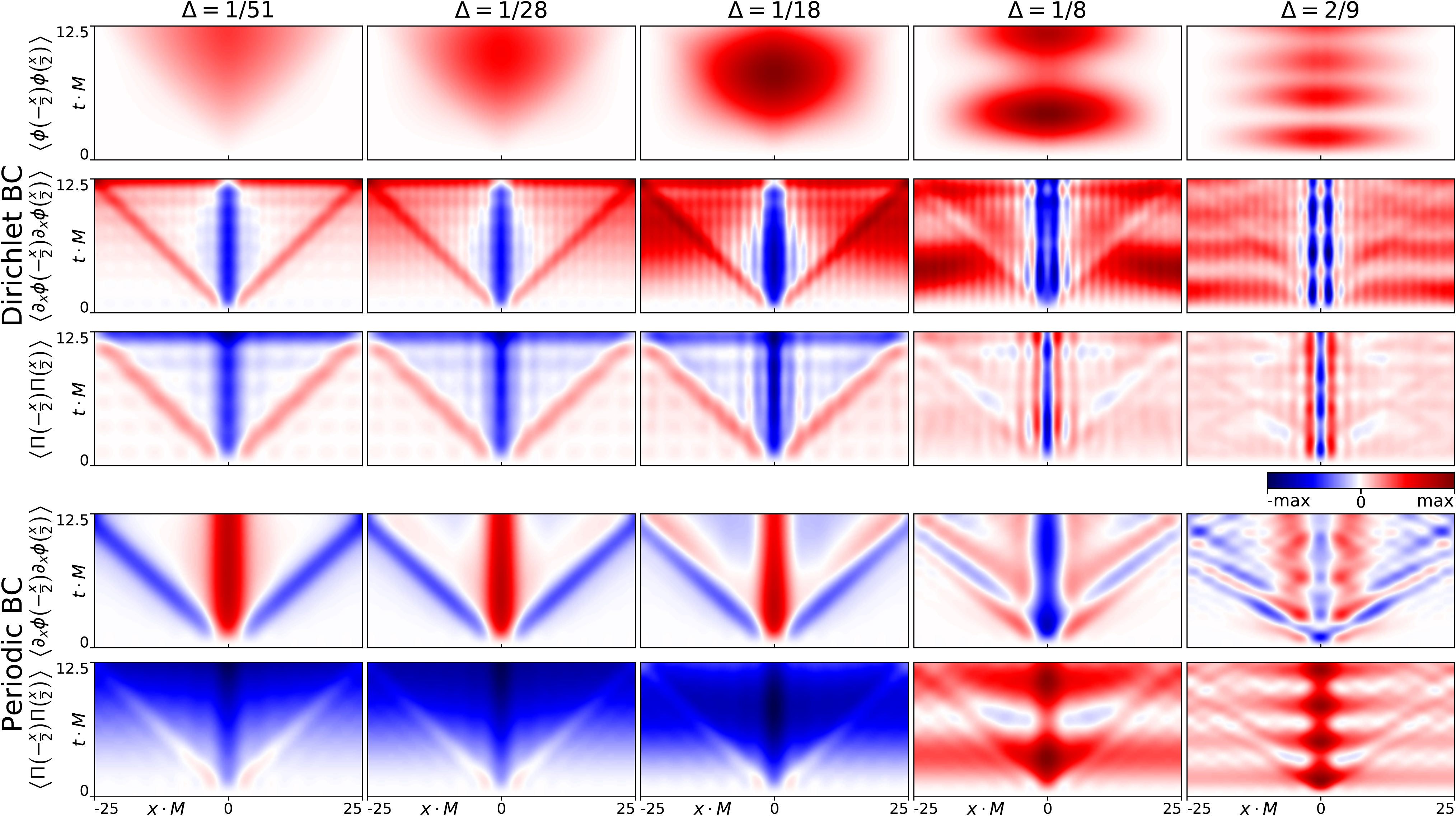}}
\caption{Density plots of connected correlations of $\Phi$, $\partial_{x}{\Phi}$ and ${\Pi}$ as functions of distance and time, obtained from TCSA simulation in a box with Dirichlet or periodic boundary conditions (DBC resp. PBC). The initial state is the KG ground state with mass $m_0$ in the DBC case or the SG ground state with first breather mass of equal magnitude $m_0=2M_0\sin\pi\Delta/(2-2\Delta)$ for PBC. The dynamics corresponds to SG with soliton mass $M=2.5\, m_0$ and increasing values of the coupling parameter $\Delta=\beta^2/8\pi$. 
Initial correlations have been subtracted from the result, and the box length is $L=10\,m_0^{-1}$. 
 \label{fig:1}}
\end{figure*}

In order to study the dynamics of correlations in the sine-Gordon model, 
we employ a recent variant \cite{KST} of a numerical
method known as ``Truncated Conformal Space Approach'' (TCSA) \cite{Yurov-Zamo} (cf. \cite{TCSA-review} for a recent review).  
This method is especially suited for the study of interacting 1+1 dimensional QFTs, it captures efficiently non-perturbative effects  
 and has recently been extended to non-equilibrium time evolution problems \cite{TCSA-QQ,KST}. 
 
The method assumes that the Hamiltonian of the model under study can be written as $H=H_{0}+\Delta H$ where $H_{0}$ is exactly solvable, while $\Delta H$ is an operator with known matrix elements
between eigenstates of $H_{0}$. Truncating the Hilbert space in finite volume $L$ by a high-energy cut-off, it becomes finite dimensional thus enabling numerical computation. 
For the SG model \eqref{eq:SG} the reference Hamiltonian $H_0$
is the massless free boson Conformal Field Theory (CFT) $H_{CFT}=\int(\tfrac{1}{2}\Pi{}^{2}+\tfrac{1}{2}(\partial_{x}\Phi)^{2})\mathrm{d}x$
and the perturbing operator is $\Delta H=- ({\mu}/{\beta})^2\int\cos\beta\Phi\,\mathrm{d}x$ 
\cite{TCSA-sG1}. For $\beta^{2}<8\pi$ the operator $\cos\beta\Phi$
is a relevant scaling operator and so the effective coupling
flows to zero at high energy, therefore the numerical results are
expected to converge with the cut-off, with a rate dependent on $\beta$; for sufficiently attractive couplings the method converges very fast. The evolution is followed for times up to $t=L/2$ to avoid boundary effects arising from the finite volume. 

Our numerical simulation for the present problem consists of the following steps:
\begin{enumerate}[topsep=1pt,noitemsep,wide,labelwidth=!,labelindent=0pt]
\item Constructing truncated matrix representations of the pre-quench and post-quench Hamiltonians;
\item Computing the initial state $|\Omega\rangle$ in the CFT basis, as the ground state of the pre-quench Hamiltonian;
\item Computing the time evolution operator $\exp(-\mathrm{i}H_{SG}t)$ by matrix exponentiation;
\item Constructing the observables $\Phi, \partial_{x}\Phi$ and $\Pi$ at different positions in space using mode expansions in the CFT basis;
\item Evaluating the correlation functions as expectation values $\langle\Omega|\mathrm{e}^{+\mathrm{i}H_{SG}t}\mathcal{O}(x)\mathcal{O}(y)\mathrm{e}^{-\mathrm{i}H_{SG}t}|\Omega\rangle$, and determining the connected
correlations $C_{\mathcal{O}}(x,y;t)$.
\end{enumerate}
We stress that the KG Hamiltonian, as well as correlations involving $\Phi$ are not generally possible to implement in TCSA due to infrared divergences from the massless boson zero mode; Dirichlet boundary conditions are an exception because the zero mode is absent.

Fig.~\ref{fig:1} shows the spreading of correlations under
SG dynamics for increasing values of the interaction $\beta$. 
In marked contrast to the expected behaviour, SG dynamics leads to strong violations of the horizon effect: for small values of $\beta$ the correlations stay within or close to the horizon, while for larger $\beta$ they quickly spread outside of the horizon, and develop temporal oscillations at a frequency increasing with $\beta$. It is most drastic for the derivative fields $\partial_{x}\Phi$ and $\partial_{t}\Phi=\Pi$, for which the out-of-horizon correlations become spatially uniform for large distance. 

We find that the violation of horizon is present universally in quenches to the SG model, independently of the initial state, as we shall discuss in Sec. \ref{sec:FurtherProperties}. The observed phenomenon has an interesting dependence on the boundary conditions: in case of Dirichlet boundary conditions (DBC) the violation is dominantly in the $C_{\partial_x\Phi}$ channel, while for periodic boundary conditions (PBC) the violation is dominated by the $C_{\Pi}$ correlations. 

\section{Analytical demonstration of the horizon violation}\label{sec:demonstration}

To validate our unexpected numerical observation and clarify the origins of the out-of-horizon effect, we perform an analytical calculation of the dynamics at a convenient value of the coupling, exploiting the powerful tool of bosonisation \cite{Delft,GNT-bosonization,Langmann2015}. Quenches in the sine-Gordon model have been studied in \cite{Iucci2010,Bertini-Essler,Cubero-Schuricht,Horvath-Takacs,Rylands2019} for a special type of initial states that contain only pairs of opposite momentum excitations. These earlier studies relied on expressing the initial state in the post-quench basis. In contrast, here we use an approach that derives directly the dynamics of correlations from initial ones \cite{Mossel2012,Kormos2014,Sotiriadis2014,Sotiriadis2017,Bastianello2017}, which in the present case relies upon the exact mapping to the massive Thirring model.

Bosonisation provides an exact non-linear and non-local correspondence between a fermionic and a bosonic QFT in 1+1 dimensions.
It relates the SG model to an interacting fermionic QFT, the massive Thirring model \cite{Coleman}, where the soliton-antisoliton fields of SG are identified with the fermion fields \cite{Mandelstam}. 
More specifically, the massive Thirring model described by the Hamiltonian
\begin{equation}
H_{MT} =\int\left[\overline{\Psi}\left(-\mathrm{i}\gamma^{1}\partial_{x}+M\right)\Psi+\tfrac{1}{2}g\left(\overline{\Psi}\gamma^{\mu}\Psi\right)\left(\overline{\Psi}\gamma_{\mu}\Psi\right)\right]\,\mathrm{d}x\nonumber
\end{equation}
where $\Psi=\left(\Psi_{-},\Psi_{+}\right)$ is a Dirac fermion field, 
is equivalent to the SG model (\ref{eq:SG}) with 
\[
\frac{\beta^{2}}{4\pi}=\frac{1}{1+g/\pi}\,,
\]
where the fermion mass $M$ is identified with the SG soliton mass. The interacting fermion fields $\Psi_\pm$ create the soliton/anti-soliton excitations and can be written as a non-local expression of the SG fields $\Phi(x)$ and $\Pi(x)$ \cite{Mandelstam}:
\begin{equation}
    \Psi_\pm(x)=\mathcal{N}:\exp\bigg[
    -\frac{2\pi \mathrm{i}}{\beta}\int\limits_{-\infty}^x \mathrm{d}\xi\,\Pi(\xi)\mp \frac{\mathrm{i}\beta}{2} \Phi(x)\bigg]:
\label{Mandelstam_vertex}\end{equation}
where $\mathcal{N}$ is a normalisation constant, and $:\,\,\,:$ denotes normal ordering. 
For $\beta^2=4\pi$ the fermion interaction $g$ vanishes
and the SG model becomes equivalent to the free massive Dirac theory
\begin{align}
H_{MF} & =\int\overline{\Psi}\left(-\mathrm{i}\gamma^{1}\partial_{x}+M\right)\Psi\,\mathrm{d}x\label{eq:MF}\,,
\end{align}
which enables analytic calculation of the time evolution.
In particular, \eqref{Mandelstam_vertex} simplifies to
\begin{equation}
\Psi_{\pm}(x)=\mathcal{N}\,:\negmedspace\mathrm{e}^{\mp\sqrt{4\pi}\mathrm{i}\mathcal{X}_{\pm}(x)}\negmedspace:\label{eq:F2B}
\end{equation}
where
\begin{equation}
\mathcal{X}_{\pm}(x)=\frac{1}{2}\left(\Phi(x)\pm\int_{-\infty}^{x}\mathrm{d}x'\,\Pi(x')\right)\,,
\label{eq:chiral_bosons}\end{equation}
with the inverse relation given by
\begin{equation}
\partial_{x}\mathcal{X}_{\pm}(x)=
\mathcal{N}':\negmedspace\Psi_{\pm}^{\dagger}(x)\Psi_{\pm}(x)\negmedspace:\label{eq:B2F}
\end{equation}

The underlying idea of our analytical solution is to take advantage of this duality by representing the initial bosonic correlations in the fermion picture, propagating them using the free Dirac dynamics and then mapping the result back to the boson picture. Although the fermionic dynamics is free, the nonlocal and nonlinear nature of bosonisation acts as a source of nontrivial physics in the problem under consideration. More precisely, our calculation consists of the following steps:
\begin{enumerate}[topsep=1pt,noitemsep,wide,labelwidth=!,labelindent=0pt]
\item Fermionising the SG model and rewriting the observables in terms of the fermionic fields using the inverse of \eqref{eq:chiral_bosons}
\begin{align}
\partial_x\Phi & = \partial_x\mathcal{X}_+ + \partial_x\mathcal{X}_- , \quad
\Pi = \partial_x\mathcal{X}_+ - \partial_x\mathcal{X}_- ,\label{eq:B2Fb}
\end{align}
together with \eqref{eq:B2F};
\item Deriving the time evolved correlations $C_{\partial\Phi}(x,y;t)$ and $C_{\Pi}(x,y;t)$ from the initial fermionic correlations  by:
\begin{enumerate}[nosep]
\item solving the free fermion time evolution 
\begin{equation}
\dot{\Psi}_{\sigma}=\sigma\partial_{x}\Psi_{\sigma}+\mathrm{i}M\Psi_{-\sigma}\nonumber
\end{equation}
in terms of initial conditions for the field as
\begin{equation}
\Psi_{\sigma}(x,t)=\sum_{\sigma'=\pm}\int\mathrm{d}x'\,G_{\sigma\sigma'}(x-x',t)\Psi_{\sigma'}(x',0)\label{eq:PropagationFermion}  
\end{equation}
where $G_{\sigma\sigma'}(x-x',t)$ is the retarded Green's function;
\item using \eqref{eq:PropagationFermion} to express the time evolution of bosonic correlations in terms of the initial fermionic correlations: 
\begin{align}
 & 
 \left\{
 {C_{\partial_x\Phi}(x,y;t) \atop C_{\Pi}(x,y;t)}
 \right\}
 ={\mathcal{N}'}^2\sum_{\sigma_i,\rho_i=\pm}
 \left\{
 {1 \atop \sigma_0\rho_0} 
 \right\}
 \int\mathrm{d}x_{1}\mathrm{d}x_{2}\mathrm{d}y_{1}\mathrm{d}y_{2}\;\nonumber \\
 & \times G_{\sigma_0\sigma_{1}}^{*}(x-x_{1},t) G_{\sigma_0\sigma_{2}}(x-x_{2},t) G_{\rho_0\rho_{1}}^{*}(y-y_{1},t) G_{\rho_0\rho_{2}}(y-y_{2},t)\nonumber \\
 & \times \Big(C_{\sigma_{1}\sigma_{2}\rho_{1}\rho_{2}}^{F}\left(x_{1},x_{2},y_{1},y_{2}\right) - C_{\sigma_{1}\sigma_{2}}^{F}\left(x_{1},x_{2}\right)C_{\rho_{1}\rho_{2}}^{F}\left(y_{1},y_{2}\right) \Big)\,,\label{eq:time-evol_corr1}
\end{align}
\end{enumerate}
\item Deriving the initial fermionic correlations 
\begin{align}
C_{\sigma\sigma'\rho\rho'}^{F}\left(x,x',y,y'\right)  &\coloneqq \left\langle\Omega\left| \,\Psi_{\sigma}^{\dagger}(x)\Psi_{\sigma'}(x')\Psi_{\rho}^{\dagger}(y)\Psi_{\rho'}(y')\right|\Omega\right\rangle \nonumber \\
C_{\sigma\sigma'}^{F}\left(x,x'\right)  &\coloneqq\left\langle\Omega\left| \,\Psi_{\sigma}^{\dagger}(x)\Psi_{\sigma'}(x') \right|\Omega\right\rangle \label{eq:initial_corr}
\end{align}
in the KG ground state $|\Omega\rangle$ as follows:
\begin{enumerate}[nosep]
    \item identifying the index combinations $\sigma$, $\sigma'$, $\rho$, $\rho'$ allowed by fermionic superselection rules;
    \item expressing fermionic correlations $C_{\sigma\sigma'\rho\rho'}^{F}\left(x,x',y,y'\right)$
in terms of bosonic ones $\langle \Omega|\mathcal{X}_\sigma(x)\mathcal{X}_{\sigma'}(x')|\Omega\rangle $ using (\ref{eq:F2B}) and exploiting the Gaussianity of $|\Omega\rangle$ in terms of the bosonic fields via Wick's theorem; 
\item obtaining correlators $\langle \Omega|\mathcal{X}_\sigma(x)\mathcal{X}_{\sigma'}(x')|\Omega\rangle $ from those of $\Phi$ and $\Pi$ using \eqref{eq:chiral_bosons}.
\end{enumerate}
\end{enumerate}

Finally we integrate numerically \eqref{eq:time-evol_corr1} to get $C_{\partial_x\Phi}(x,y;t)$ and $C_{\Pi}(x,y;t)$ for finite $r=|x-y|$, and compute analytically the asymptotics at large $r$ for any time $t$ in App.~\ref{app:asympt} to verify the horizon violation effect explicitly.

\begin{figure*}[!ht]
\centering
\includegraphics[clip,width=0.46\textwidth]{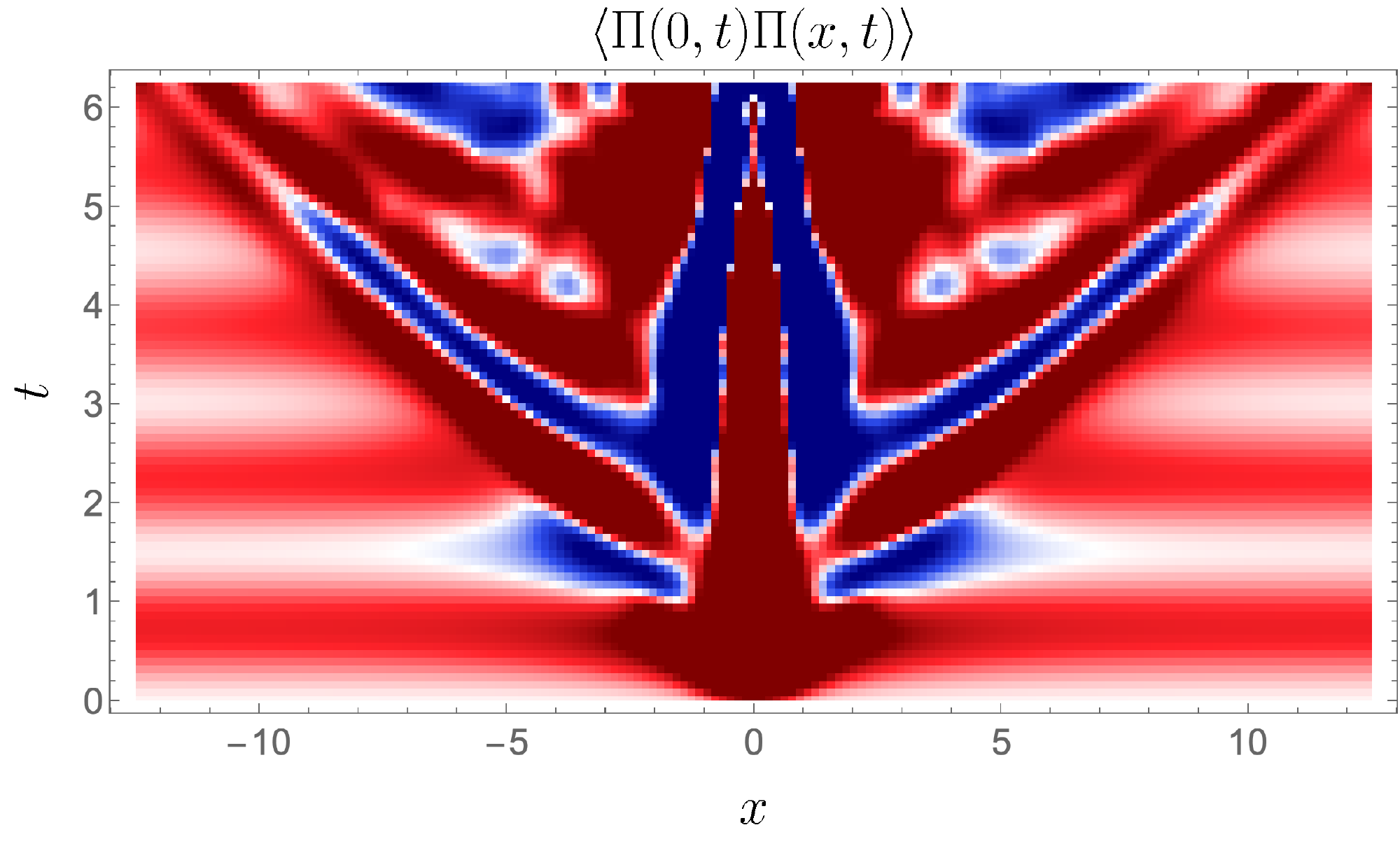} 
\includegraphics[clip,width=0.03\textwidth]{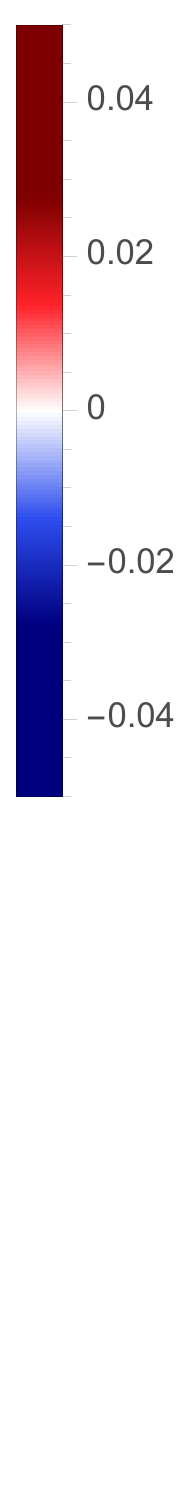} 
\includegraphics[clip,width=0.45\textwidth]{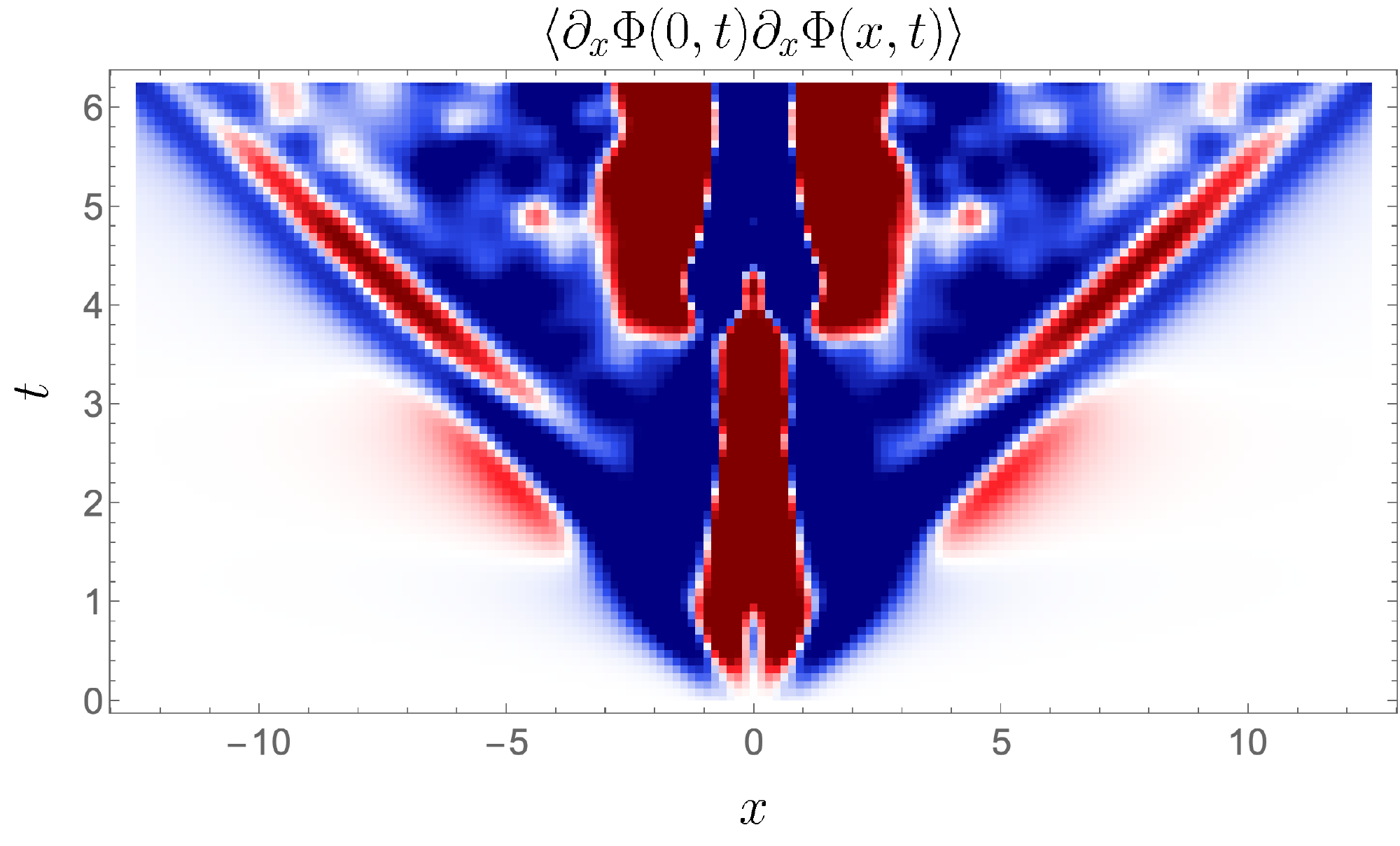}
\includegraphics[clip,width=0.03\textwidth]{legend.pdf} 
\caption{The analytical results for the correlations of $\Pi$ and $\partial_x\Phi$ under SG dynamics at $\beta=\sqrt{4\pi}$ for PBC. 
Density plot of $\Pi$ (\emph{top}) and $\partial_x\Phi$ (\emph{bottom}) correlations as function of distance $x$ and time $t$ (in units $1/M$). Initial correlations have been subtracted as in Fig.~\ref{fig:1}. The initial state is the KG ground state with mass $m_0$ and the post-quench soliton mass is $M=2.5m_0$.
\label{fig:2a}}
\end{figure*}
\begin{figure}[!ht]
\centering
\includegraphics[clip,width=.6\columnwidth]{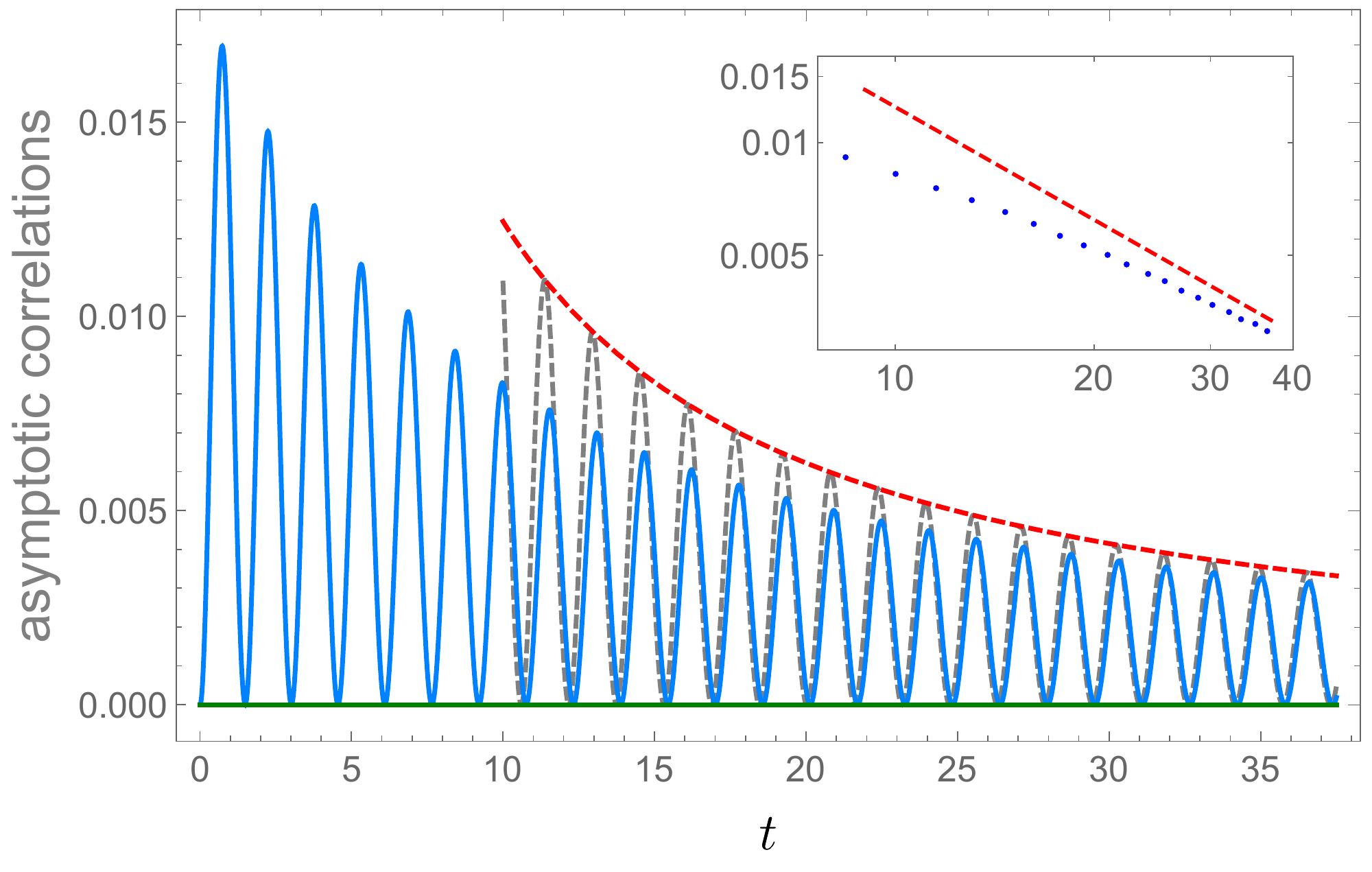} 
\caption{
Time evolution of the large distance asymptotics $\lim_{x\to\infty}\langle\Pi(0,t)\Pi(x,t)\rangle$  (blue line) together with their large time scaling $\sim\sin^2(2Mt)/t$ and envelope (gray and red dashed lines respectively). The vanishing large distance asymptotics of $\partial_x\Phi$ correlations (green line) are included for completeness. \emph{Inset}: log-log plot showing the large time scaling of asymptotic $\Pi$ correlations. The  points are the maxima of the oscillations and the red dashed line the $\sim 1/t$ envelope. The quench parameters are same as in Fig.~\ref{fig:2a}.
\label{fig:2b}}
\end{figure}
The resulting exact time evolution of $C_{\Pi}$ and $C_{\partial_x\Phi}$ is shown in Fig.~\ref{fig:2a} and \ref{fig:2b}. The analytical results confirm the dynamical emergence of out-of-horizon connected correlations, which are only present for the field $\Pi(x)$ and not for $\partial_x\Phi(x)$, consistently with the numerical TCSA results. Note that the horizon violation persists up to infinite distances and its long range asymptotics exhibits oscillations algebraically decaying with time, as shown in Fig.~\ref{fig:2b}.

\section{Physical explanation of the effect}\label{sec:explanation}

The analytical solution by bosonisation not only confirms the observed horizon violation but also provides a transparent explanation for the mechanism of the observed phenomenon. The correlation functions $C_{\partial_x\Phi}(x,y;t)$ and $C_{\Pi}(x,y;t)$ can be expressed as quadruple convolutions of the free fermion propagator $G_{\sigma\rho}(x-x',t)$ (cf. eqs.~\eqref{eq:time-evol_corr1} and \eqref{eq:initial_corr}) with the initial four-point correlations of the fermion fields on the KG ground state $|\Omega\rangle$. The fermion propagators vanish identically outside of the light-cone ($|x-x'|>t$) so they cannot be the source of the horizon violation, i.e. the dynamics is strictly causal in the relativistic sense.

The initial fermionic correlations, however, display unexpected behaviour which is where the observed effect originates from. Among the six terms allowed by the fermionic superselection rules, two of them,
\begin{equation}
    C_{\pm\mp\mp\pm}^{F}\coloneqq\langle\Omega|\Psi_\pm^\dagger(x)\Psi_\mp(x+a)\Psi_\mp^\dagger(y)\Psi_\pm(y+b)|\Omega\rangle
\end{equation}
violate the cluster decomposition principle. That is, these terms do not factorise to a product of two-point functions asymptotically as $|x-y|\rightarrow\infty$. More precisely, $C_{\pm\mp\mp\pm}^{F}$ tends to a nonzero value in this limit while the two-point functions vanish. Thus, the ground state of the KG model $|\Omega\rangle$ clusters perfectly in terms of local bosonic fields but does not satisfy clustering for the fermionic fields which form the natural basis for the post-quench time evolution.

\begin{figure*}[!ht]\
	\centering
	\includegraphics[clip,width=.8\textwidth]{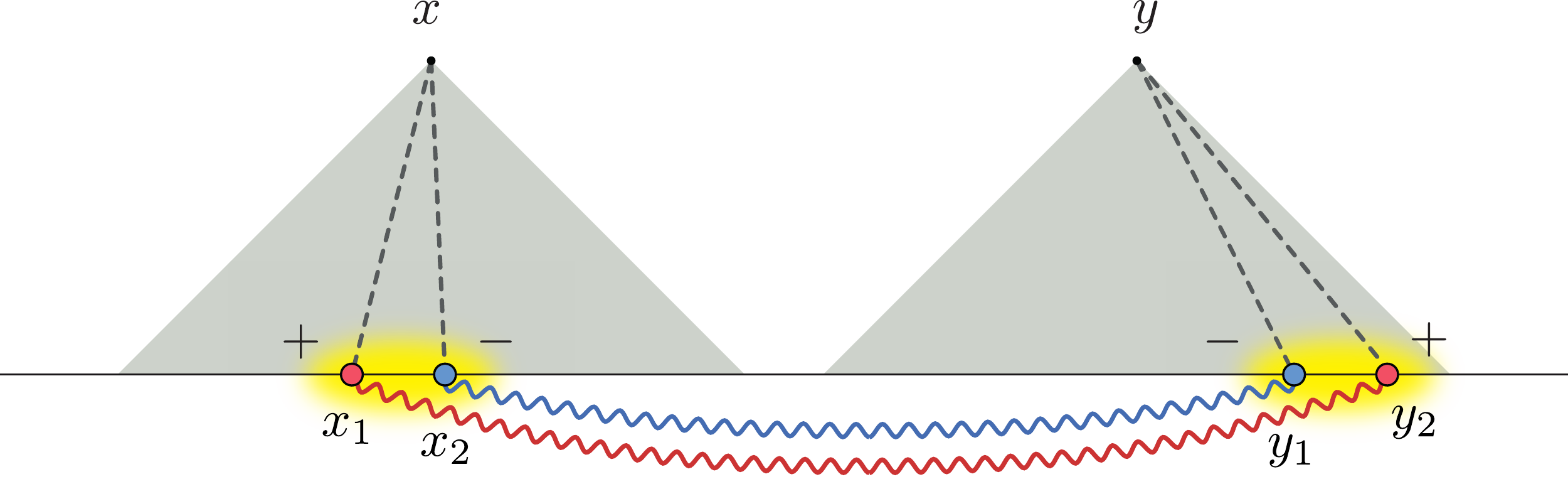}\caption{Graphical explanation of how a non-zero out-of-horizon correlation $C_{\Pi}(x,y;t)$ emerges due to the presence of long-range initial fermionic correlations. The coloured links depict the correlation $C_{+--+}^{F}(x_1,x_2,y_1,y_2)$, while the black dashed lines are the fermion (i.e. soliton) propagators. The yellow regions show the correlation length $m_0^{-1}$ of the pre-quench KG model.\label{fig:3}}
\end{figure*}

Note that such correlations are undetectable at $t\leq 0$ by measurements of local bosonic observables in the initial state. However for $t>0$, the long-range correlations of the fermionic  degrees of freedom get dynamically revealed even in the correlations of local fields, and dominate the asymptotic correlations of $\Pi$. 

Our analytical solution explains the mechanism of the horizon violation effect at the free fermion point and suggests that there, the effect is a consequence of a four-body process involving two soliton - antisoliton pairs. This manifests itself in the frequency of out-of-horizon oscillations, which is $4M$ i.e. four times the soliton mass. This picture is expected to stay valid beyond the free fermion point into the repulsive regime $\beta^2>4\pi$. However, in the attractive regime a different behaviour is expected due to the presence of bound states.

\begin{figure}[ht!]
	\centering{\includegraphics[clip,width=.7\columnwidth]{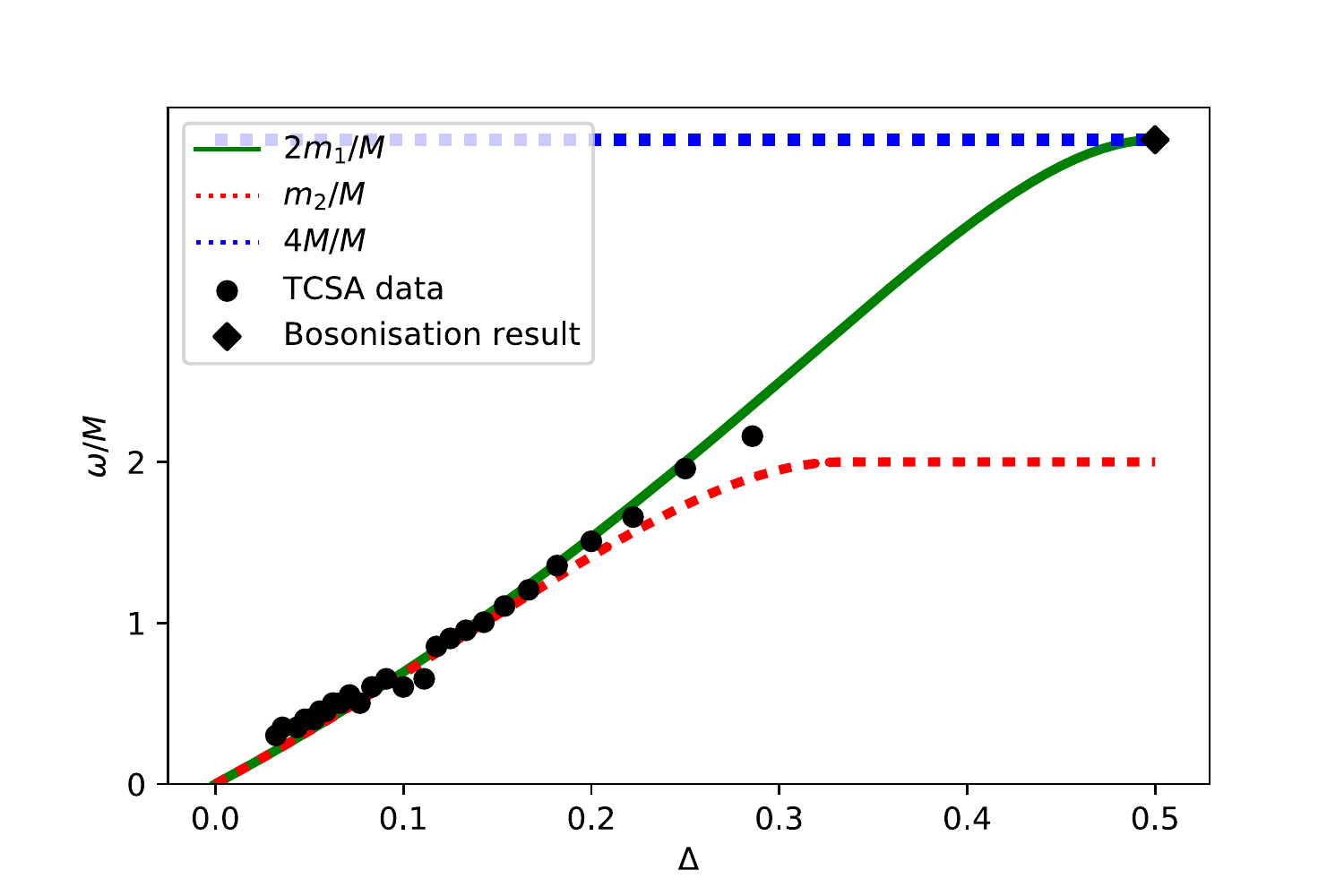}
		\caption{Oscillation frequencies of the horizon violation effect. Numerical data from the TCSA and the analytical result at the free fermion point using bosonisation, are compared with several relevant masses present in the theory: twice the mass of the first breather $2m_1$, mass of the second breather $m_2$ and four times the soliton mass $4M$. The TCSA works best in the small and intermediate $\Delta$ regime and becomes progressively less convergent for $\Delta\gtrsim0.3$.
			\label{fig:frequencies}}}
\end{figure}

This is verified by an analysis of the oscillation frequency based on the TCSA data for quenches in the attractive regime, as shown in Fig.~\ref{fig:frequencies}. The data suggest that the oscillation frequency in this case is $2m_1$, i.e. twice the mass of the lightest breather $B_1$ which is a bound state of a soliton and an anti-soliton. Note that this frequency changes continuously into $4M$ at the free fermion point, consistently with the analytical result. This suggests that in the attractive regime, the horizon violation is predominantly mediated by pairs of $B_1$ breathers, instead of unbound soliton - antisoliton pairs. Such an interpretation leads to a prediction: In the free fermion case and in the repulsive regime, the continuous spectrum of the soliton - antisoliton pairs leads to dephasing and thus to the  time-decay of the out-of-horizon component as observed in the analytical calculation. In the attractive regime on the other hand, this dephasing effect is expected to be suppressed due to the fact that breathers are isolated bound state excitations below the two-particle continuum. We therefore expect that there should be no decay of the out-of-horizon component or it should be significantly slower. This expectation seems to be consistent with the TCSA plots, even though our observation of the effect is limited by the finite system size. 

Solitons are topological particles emerging because the boson field is compactified to a circle as imposed by the periodic potential of the SG model. This results in infinitely many vacua characterised by the topological charge (a.k.a. winding number), and solitons interpolating between them appear as the fundamental topologically charged particles. Since the clustering violation appears in the initial state correlations of the soliton creating fields, this suggests that the horizon violation is due to their topological nature. The role of topology is further supported by the fact that similar clustering violation has also been found in ground states of theories with gauge field configurations characterised by a non-trivial Pontryagin number \cite{2d-QFT}.

\section{Further properties of the horizon violation}\label{sec:FurtherProperties}

\subsection{Universal presence in quenches to the sine-Gordon model}

\begin{figure}[ht!]
	\centering{\includegraphics[clip,width=\columnwidth]{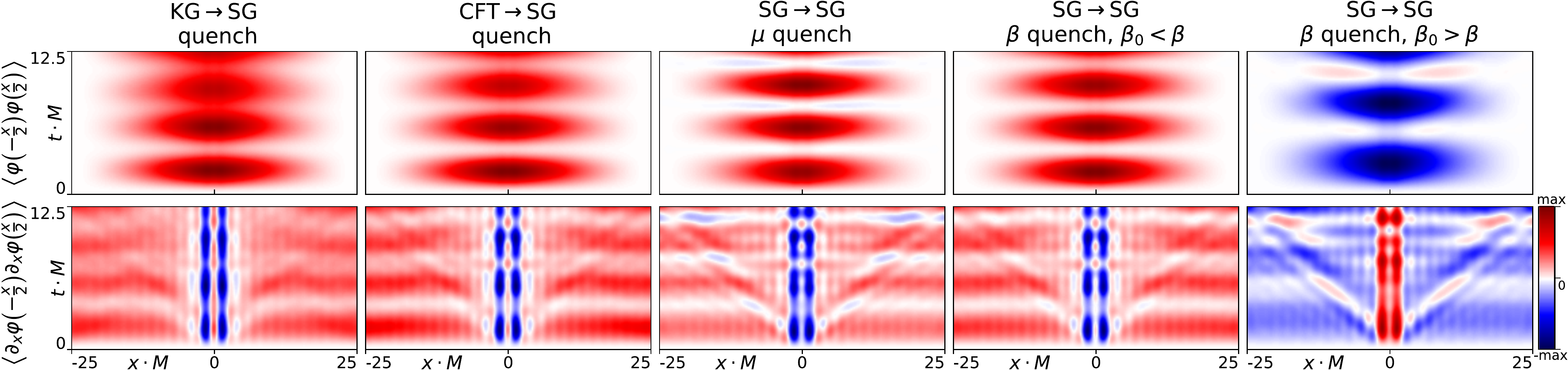}
		\caption{Density plots of correlation dynamics for different types of quenches: from the KG ground state, the massless free boson (CFT) ground state, SG mass $\mu$ quench, SG interaction quench from smaller to larger interaction $\beta$ or vice versa. Plots are for Dirichlet boundary conditions where it is technically easier to perform a TCSA simulation for all these different types of quench.
			\label{fig:SM1a}}}
\end{figure}

In order to clarify whether the horizon violation is a special property of the particular choice of the Klein-Gordon ground state as the pre-quench state, we perform additional numerical simulations studying quenches from a variety of other physical initial states. In the absence of analytic solutions we have to rely on the numerical results. We find that the horizon violation is not restricted to quenches from Klein-Gordon to sine-Gordon but is universally present in quenches to the sine-Gordon model as demonstrated in Fig.~\ref{fig:SM1a}. The violation was observed in quenches from the massless free boson (CFT) and in SG to SG quenches both in mass $\mu$ and interaction $\beta$. The interpretation is that any quench to the SG model or any global change of a parameter correlates pairs of soliton -- anti-soliton pairs resulting in infinite range correlations.

Some details of the effect do change for different initial states, such as the overall sign of the correlations if the quench is from a higher value of the $\mu$ or $\beta$ parameter to a lower one as compared to the opposite direction. This sign flip is quite generic: it is also present in free KG to KG quenches which show no horizon violation. Another interesting detail is the dependence on boundary conditions, discussed below.

\subsection{Sensitivity to boundary conditions}

The horizon violation has an interesting sensitivity to boundary conditions as shown in Fig.~\ref{fig:1}. For Dirichlet boundary conditions the violation is dominated by correlations of the $\partial_x\Phi$ field, while for periodic boundary conditions it switches over to $\Pi$. This sensitivity to boundary conditions is another sign of the topological origin of the phenomenon.

In Fig.~\ref{fig:1}, the quenches with Dirichlet boundary conditions (Dirichlet BC, DBC) are started from a massive KG ground state which is the canonical example of a state with exponentially decaying correlations and satisfying clustering. This is also the initial state that we use in the analytical solution based on bosonisation. The pre-quench KG mass $m_0$ is chosen so that the initial correlation length is 1/10 of the system size, $\xi_0/L=1/10$. For periodic boundary conditions (periodic BC, PBC), however, it is difficult to implement the KG ground state in the TCSA Hilbert space of the SG model due to the infrared divergent zero mode of the boson field, therefore for PBC we study SG to SG quenches in $\mu$. In SG ground states the correlation length is controlled by the mass $m_1$ of the lightest particle, i.e. the first breather, which depends on the soliton mass $M$ and the interaction $\Delta=\beta^2/8\pi$ as $m_1=2M\sin(\pi\Delta/(2-2\Delta))$. The soliton mass itself can be related to the Hamiltonian parameters using Zamolodchikov's mass -- coupling relation \cite{Zamo1995}. We choose the pre-quench $\mu_0$ so that $m_1L=10$ and we thus again have $\xi_0/L=1/10$. Note that due to the zero mode the field $\Phi$ is not unambiguously defined for periodic BC and so we only compute correlations of the  derivative fields ($\partial_x\Phi$ and $\Pi=\partial_t\Phi$ from the TCSA.

The features of the horizon violation observed for PBC agree with our analytical solution which was also performed for PBC, however starting from a KG ground state, since the ground state of the interacting SG model is not accessible with state of the art analytic techniques. As we have discussed, the features of the effect do not depend qualitatively on the choice of the pre-quench state.

\section{Prospective realisation in ultra-cold atom experiments}\label{sec:experiment}

Experimental techniques in ultra-cold atoms have rapidly developed in the last two decades and today a broad variety of quantum many-body Hamiltonians can be implemented enabling the study of their equilibrium and out-of-equilibrium physics. A type of ultra-cold atom experiments particularly suitable for the study of quantum field theory are experiments in atom chips \cite{SchmiedmayerChapter2018,LangenThesis2015,exp:rev-3,exp-GGE,exp-sG,LCexp2,phase_locking,Pigneur2018,SchmiedmayerRecurrences2018}. Using state of the art magnetic traps, strong 1D trapping potentials can be generated which confine $\,^{87}\text{Rb}$ atomic quasi-condensates of around 5000 atoms to 1D elongated geometries. Although consisting from individual atoms, on scales longer than the so-called healing length $\xi_h$ such 1D quasi-condensates effectively behave as a continuum and their physics is described by quantum field theories. Individual 1D condensates are realisations of the Luttinger liquid (free Bose gas), while coupling two such condensates via Josephson tunneling results in a system governed by sine-Gordon dynamics \cite{Gritsev2007b}. More precisely, the SG model describes the dynamics of relative phase between the two condensates, the SG coupling parameter $\beta$ is related to the Luttinger parameter $K$ of the individual quasi-condensates and the SG mass parameter $\mu$ is related to the Josephson frequency $J$.

Atom chip setups have been used to study nonequilibrium dynamics of quantum field theories following quantum quenches \cite{exp-GGE,LCexp2,phase_locking,SchmiedmayerRecurrences2018} and to measure directly for the first time higher order correlations of an interacting quantum field theory (the SG model) \cite{exp-sG}. Therefore, all the necessary techniques needed to experimentally measure the horizon violation presented in this manuscript have already been developed and successfully applied in several contexts, making the effect accessible to atom chip experiments.

A possible protocol to measure horizon violation in atom chip experiments can be obtained exploiting the fact that the phenomenon is present also in the SG to SG quenches, in which case it could be realised in the following steps:
\begin{enumerate}
	\item A system of two coupled 1D quasi-condensates with Josephson frequency $J_0$ is prepared in a low-temperature thermal state. Such a state corresponds to a thermal state of the SG model with the mass $\mu_0$.
	\item  The potential barrier separating the two quasi-condensates is changed abruptly which quenches the Josephson frequency to a new value $J$. This corresponds to the SG to SG mass quench from $\mu_0$ to $\mu$.
	\item The system is left to evolve in time and the equal-time two-point correlations of the relative phase and the density fluctuations are measured. These correspond to the $C_{\Phi}$ and $C_{\Pi}$ and one can use the former to extract $C_{\partial_x\Phi}$.
\end{enumerate}
In case such a protocol is realised, the prediction is that the correlations of the relative phase and relative density fluctuation must show a tail not decaying with distance and oscillating in time.

\section{Conclusions}\label{sec:conclusions}

{Using the truncated conformal space approach we demonstrated that the connected correlations following quenches in the sine-Gordon model appear outside of the horizon, with an out-of-horizon component of the correlations which does not decay with distance and oscillates in time. This is the first explicit counter example to the horizon effect, which is a feature of time evolution in non-equilibrium quantum field theory that so far has been expected to be generally valid. For the special case of quenches from Klein-Gordon to sine-Gordon model at the free fermion point, we established the horizon violation analytically using bosonisation.} 

{We showed that the violation of horizon is a consequence of an interesting property of quenches to the SG model: even though the initial states (ground states of the Klein-Gordon and the SG models at different parameter values) have exponentially decaying correlations and cluster in terms of the local bosonic field, they violate clustering in terms of solitonic fields which govern the post-quench dynamics of the SG model. Before the quench this clustering violation is undetectable by local measurements of bosonic observables, but appears in the correlations of local bosonic observables after the quench due to the nonlinear dynamics of the SG model. We have also proposed a protocol to measure the phenomenon in experiments with ultra-cold atoms in atom chips.}

{Let us now turn to the physical interpretation of the horizon violation. Firstly, it is clear that there is no violation of causality involved. Indeed, the sine-Gordon model is relativistically invariant satisfying micro-causality i.e. all commutators between space-like separated observables vanish \cite{Wightman,Streater-Wightman,Haag,Ruelle}. In the analytic bosonisation approach this is manifest since the propagator used in the computation vanishes outside of the past light-cone. The effect thus completely originates from the global quantum quench performed at $t=0$.}

{It is known that non-vanishing connected correlations exist between space-like separated regions in the vacuum (and more generally in a dense set of states in the Hilbert space) of any relativistic quantum field theory \cite{Wald,Witten,Reznik2003,ReznikRetzker2005}, as a consequence of the Reeh-Schlieder theorem \cite{Reeh}. However, in a theory with a non-vanishing mass gap and a unique vacuum \cite{Araki}} such correlations satisfy clustering with their connected component decaying exponentially with distance, and so cannot explain the observed effect.

The origin of the horizon violation in the non-equilibrium time evolution is that initial correlations involving soliton creating fields do not satisfy clustering. This is due to the soliton fields being non-local expressions in terms of the local bosonic field, as demonstrated explicitly by bosonisation. Such clustering violation is known to occur in ground states of theories with gauge field configurations characterised by a non-trivial Pontryagin number \cite{2d-QFT}. Together with the dependence on boundary conditions, these considerations strongly support the topological origin of the horizon violation effect.

Note that the system after the quench is in a highly-excited non-equilibrium state of the post-quench Hamiltonian, therefore the appearance of clustering violation does not contradict any existing theorems of quantum field theory, and it can eventually be considered a transient feature of the non-equilibrium dynamics.

{Our results show that the horizon effect which has been widely accepted as a general feature of dynamics in non-equilibrium QFT is not generally valid: quantum quenches can generate infinite range correlations in a class of quantum field theories with non-trivial field topology.}

{It would be interesting to investigate the horizon violation from the quantum information point of view. From the structure of the initial correlations, it is natural to expect that the effect is a quantum field theoretic version of long-distance entanglement of the Einstein-Podolsky-Rosen (EPR) type \cite{EPR}: the asymptotic nonvanishing of $C_{\pm\mp\mp\pm}^{F}$ could be a manifestation of long-distance entanglement between soliton -- anti-soliton pairs, as depicted in Fig.~\ref{fig:3}. Research in this direction is left for future work.}

\begin{acknowledgments}
This work was partially supported by the Advanced Grant of European
Research Council (ERC) 694544 -- OMNES, by the Slovenian Research
Agency under grants N1-0055 (OTKA-ARRS joint grant) and P1-0402, as well as by the
National Research Development and Innovation Office of Hungary within
the Quantum Technology National Excellence Program (Project No. 2017-1.2.1-NKP-2017-00001) and under grants OTKA No.~SNN118028 
and K-16 No.~119204. The work of I.K. was also supported by the Max-Planck-Harvard Research Center for Quantum Optics (MPHQ). S.S. acknowledges support by the Slovenian Research
Agency under grant N1-0109 (QTE). 
The work of G.T. was also partially supported 
by the BME-Nanotechnology FIKP grant of EMMI (BME FIKP-NAT). 
I.K. is grateful to Martin Horvat for support in high-performance computing.
\end{acknowledgments}

\bibliographystyle{utphys}
\bibliography{SG_LC}

\clearpage

\appendix

\section{Horizon effect for free dynamics and local initial states\label{sec:Horizon_freeQFT}}

\begin{figure}[ht]
\centering{\includegraphics[clip,width=0.7\columnwidth]{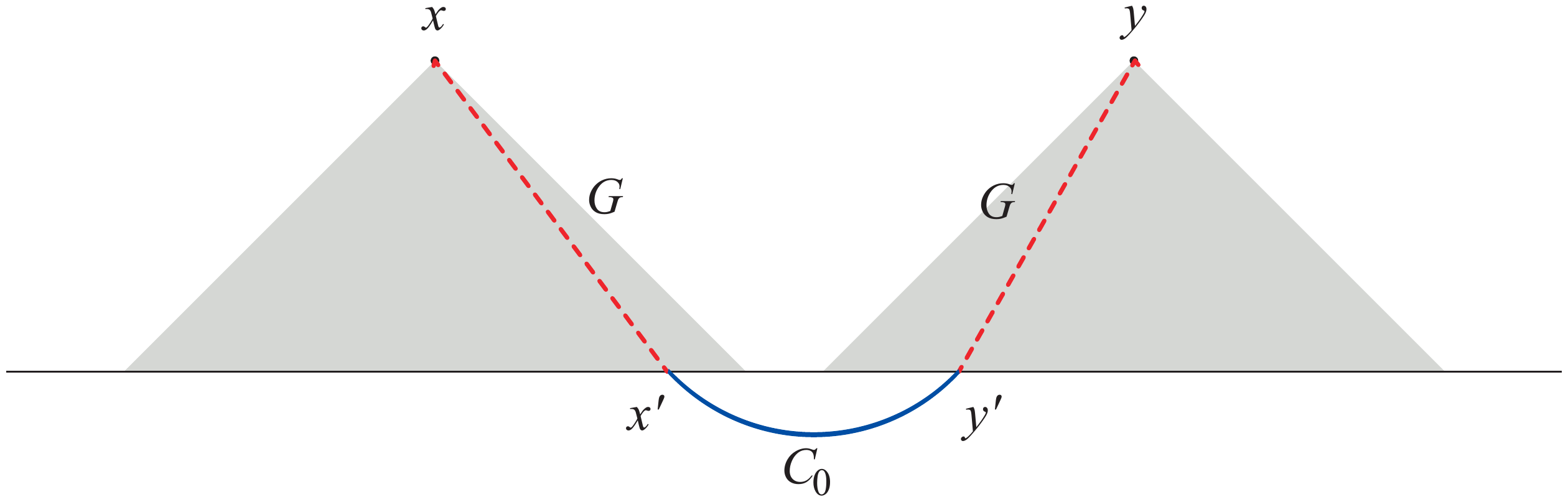}
\caption{Schematic explanation of the horizon effect \label{fig:SM0}}}
\end{figure}

Here we show that the horizon effect is always present in the special
case of free relativistic dynamics for any initial state $|\Omega\rangle$
that exhibits exponential clustering of correlations. More precisely
the condition is that the dynamics is free in terms of some choice
of local fields and the initial state satisfies exponential clustering
in terms of the same fields. The horizon effect can be stated mathematically
as follows
\begin{equation}
\left|C_{O}(x,y;t)\right|=\left|\langle\Omega|{\mathcal{O}}(x,t){\mathcal{O}}(y,t)|\Omega\rangle-\langle\Omega|{\mathcal{O}}(x,t)|\Omega\rangle\langle\Omega|{\mathcal{O}}(y,t)|\Omega\rangle\right|<A\,\mathrm{e}^{-\left(|x-y|-2t\right)/\xi_{h}}\label{eq:horizon}
\end{equation}
where the length $\xi_{h}$ can be called `horizon thickness', $A$
is independent of $x$ and $y$ and as usual we have set the speed
of light equal to unit $c=1$. To demonstrate this relation we
focus on the example of Klein-Gordon dynamics, on translationally
invariant initial states and choose as local observable the field
${\Phi}$ itself, even though the reasoning holds more generally.
Because the Hamiltonian that describes the dynamics is free, the Heisenberg
equations of motion are linear which means that they can be solved
for general initial conditions using Green's functions
\begin{align}
{\Phi}(x,t) & =\int\mathrm{d}x'\left(\partial_{t}G(x-x',t){\Phi}(x')+G(x-x',t){\Pi}(x')\right)\nonumber \\
 & \equiv\sum_{i=0,1}\int\mathrm{d}x'G_{i}(x-x',t){\phi}_{i}(x')\label{eq:1}
\end{align}
where we denote ${\phi}_{0}={\Phi},{\phi}_{1}={\Pi}$,
$G_{0}=\partial_{t}G,G_{1}=G$ and the Green's function is
\begin{equation}
G(x-x',t)=\int\frac{\mathrm{d}k}{2\pi}\,\mathrm{e}^{\mathrm{i}k(x-x')}\frac{\sin E_{k}t}{E_{k}}\label{eq:2}
\end{equation}
with $E_{k}=\sqrt{k^{2}+m^{2}}$. Therefore one obtains 
\begin{align}
C_{\Phi}(x,y;t) & =\sum_{i,j=0,1}\int\mathrm{d}x'\mathrm{d}y'\,G_{i}(x-x',t)G_{j}(y-y',t)C_{\phi_{i},\phi_{j}}(x',y';0)\label{eq:3}
\end{align}
which can be depicted schematically as shown in Fig.~\ref{fig:SM0}.

Since the dynamics corresponds to a local relativistically invariant
theory, the commutators $[{\Phi}(x,t),{\phi}_{i}(x',0)]$
vanish outside of the past light-cone i.e. for $|x-x'|>t\ge0$, which
means equivalently that the retarded Green's function $G(x-x',t)$
has support only in the interval $x'\in[x-t,x+t]$.
This can be easily verified from (\ref{eq:2}) by application of Cauchy's
theorem and noticing that the integrand is an analytic function of
$k$ in the complex $k-$plane and decays exponentially in the upper
or lower half $k-$plane for $x-x'>t$ or $x-x'<-t$ respectively.
Because the initial state satisfies exponential clustering for local
field correlations we have
\begin{align*}
\left|C_{\phi_{i},\phi_{j}}(x',y';0)\right| & =\left|\langle\Omega|{\phi}_{i}(x'){\phi}_{j}(y')|\Omega\rangle-\langle\Omega|{\phi}_{i}(x')|\Omega\rangle\langle\Omega|{\phi}_{j}(y')|\Omega\rangle\right|<c_{ij}\mathrm{e}^{-|x'-y'|/\xi_{0}}
\end{align*}
where $c_{ij}$ is some constant and $\xi_{0}$ is the correlation
length characterising the initial state. Substituting in (\ref{eq:3})
and taking into account the support of the functions $G_{i}$ we deduce
that
\[
\left|C_{\Phi}(x,y;t)\right|<A\,\mathrm{e}^{-\left(|x-y|-2t\right)/\xi_{0}}
\]
where $A=\displaystyle\sum_{i,j=0,1}c_{ij}\int\mathrm{d}x'\,\left|G_{i}(x',t)\right|\int\mathrm{d}y'\left|G_{j}(y',t)\right|$
is independent of $x$ and $y$. 

This proves (\ref{eq:horizon}) and shows that for the free case the horizon thickness $\xi_{h}$
is equal to the initial correlation length $\xi_{0}$.

\section{Details of the TCSA simulations}\label{app:TCSA}

As explained in the main text, the TCSA simulations are based on a suitable truncation of the Hilbert space, so that the computation is performed by operations on finite size matrix representations of the Hamiltonians, states and observables. The CFT eigenstate basis is used as basis of the truncated Hilbert space and an energy cutoff is used for the truncation (we refer the reader to \cite{KST} for a detailed presentation of the method). The quality of the computed results is confirmed by analysing the convergence of the data for increasing values of the truncation cutoff. For the purposes of the present study we used TCSA bases of dimension ~20000 to 28000 states. The rate of convergence was fast enough for all parameter values used in the present work, and we checked that the results have converged sufficiently well for the above values of truncated basis dimensions (cf. also \cite{KST}). 

The high energy cut-off leads to truncation errors in the computed energy spectra and observables. These errors decrease when the cut-off is increased and the rate of convergence depends on the parameters of the quench: it is better for smaller quenches and for smaller values of the post-quench interaction parameter $\Delta$. Truncation errors generally increase from left to right in the sequence of density plots of Fig.~\ref{fig:1}. 
They also depend on the choice of observable and are stronger for $\partial_x\Phi$ and $\Pi$ than for $\Phi$, due to the derivative enhancing the contribution of large wave-number modes. Moreover they are stronger at short distances (i.e. in a narrow central part of the density plots) due to the presence of universal short-distance singularities, however this is not relevant for the horizon effect and its violation which take place at intermediate to large distances. 
In the present computation the maximum wave-numbers $k_\text{cut}$ used (which determine the finest spatial resolution $L/k_\text{cut}$ that can be achieved) range between 28 and 30 in the DBC case and between 11 and 12 in the PBC case, for the same order of truncated space dimension. 
The truncation errors in the spectrum also lead to the unitary time evolution getting out of phase at larger times, since it is governed by frequencies determined by the energy level differences. However, in this work we are interested in short time scales $t<L/2$, where this effect does not play an important role.

\section{Analytical derivation of horizon violation in the SG model}\label{app:BF}

In this section we give more details about the analytical solution of the quench from the Klein-Gordon model to the sine-Gordon model that we presented in the main text. We follow the three conceptual steps outlined in the main text:
\begin{enumerate}
    \item Fermionising the sine-Gordon model,
    \item Computing the dynamics of the correlation functions in terms of initial fermionic correlations,
    \item Deriving the initial fermionic correlations from the bosonic ones.
\end{enumerate}
Finally we discuss how to extract exact asymptotic expressions for the connected correlations functions at large distances and clarify in this way the origin of the observed out-of-horizon effect.

\subsection{Fermionising the sine-Gordon model}

Due to strong coupling, the dynamics of the sine-Gordon theory cannot be accessed perturbatively. The model is integrable but the current state of the art methods of the theory of integrability do not allow for computation of dynamical multi-point correlation functions. Our solution therefore relies on a powerful analytical tool, the theory of bosonisation. 

Bosonisation is an example of QFT dualities, which allow studying strongly interacting QFTs. The idea is to introduce a nonlinear field transformation, such that the original strongly-interacting model is mapped into a weakly or non-interacting model in terms of the new fields. In modern theoretical physics, dualities play a central role in understanding the physics of quantum fields and unveiling the deeper symmetries of Nature. Bosonisation establishes a mapping between two different (1+1)-dimensional QFTs, one of which is bosonic and the other fermionic. This is achieved through an isomorphism between the Hilbert spaces and the operators of the theories. This isomorphism can be rigorously established for a finite system of size $L$ \cite{Delft}; however, one can take the thermodynamic limit $L\rightarrow\infty$ in closed form.

The sine-Gordon model is a bosonic theory governed by the Hamiltonian
\begin{equation}
H_{SG}=\int\left(\frac{1}{2}{\Pi}{}^{2}+\frac{1}{2}(\partial_{x}\Phi)^{2}-{ \frac{\mu^2}{\beta^2}}\cos\beta\Phi\right)\mathrm{d}x\label{eq:SGapp}
\end{equation}
which can be mapped to a fermionic theory called the massive Thirring model \cite{Coleman,Mandelstam,Pogrebkov1975}:
\begin{equation}
H_{MT} =\int\left[\overline{\Psi}\left(-\mathrm{i}\gamma^{1}\partial_{x}+M\right)\Psi+\tfrac{1}{2}g\left(\overline{\Psi}\gamma^{\mu}\Psi\right)\left(\overline{\Psi}\gamma_{\mu}\Psi\right)\right]\,\mathrm{d}x\label{eq:MassiveThirringapp}
\end{equation}
The relation between the couplings is given by \cite{Coleman}:
\[
\frac{\beta^{2}}{4\pi}=\frac{1}{1+g/\pi}\,,
\]
At the so-called "free fermion point" $\beta=\sqrt{4\pi}$ the Thirring interaction vanishes and the dual theory of the SG model is thus the theory of a free Dirac fermion field governed by the Dirac Hamiltonian:
\begin{align}
H_{MF} & =\int\overline{\Psi}\left(-\mathrm{i}\gamma^{1}\partial_{x}+M\right)\Psi\,\mathrm{d}x\label{eq:MFapp}\,,
\end{align}
for which the time evolution dynamics can be calculated analytically.

\subsubsection{Equivalence of Hilbert spaces}  
Consider a fermionic field theory described by a set of momentum modes $c_{k,\sigma}$, $k\in\mathbb{Z},\sigma=\pm$, satisfying  canonical anti-commutation relations. Further, we assume that the vacuum of the theory is given by the Fermi sea for which all momentum modes with $k\leq0$ are occupied, while the ones with $k>0$ are empty. Then any excitation on top of the Fermi sea can be decomposed into a part that merely changes the expectation value of the fermionic number operator and a part containing only particle-hole excitations. The particle-hole excitation operators:
\begin{equation}
    a_{k}^{\dagger}=\frac{\rm i}{\sqrt{|n_{k}|}}\sum_{n_q=-\infty}^{\infty}c_{q+|k|,\sigma}^{\dagger}c_{q,\sigma}\quad \text{with }\sigma=-\text{sign}(k)\label{eq:AprticleHolesBosons}
\end{equation}
have all the algebraic properties of bosonic excitation operators. The fermionic Hilbert space can thus be decomposed as:
\begin{equation}
    \mathcal{H}_{\text{Fermi}}=\mathcal{H}_{{N}_{-},{N}_{+}}\otimes\mathcal{H}_{\text{Bose}}\label{eq:HilbertSpaceDecomposition}
\end{equation}
where the states in $\mathcal{H}_{{N}_{-},{N}_{+}}$ correspond to sectors with  different expectation values of the number operator ${N}_{\sigma}$ and $\mathcal{H}_{\text{Bose}}$ is spanned by all possible particle-hole excitations. The bosonic character of the particle-hole excitations also enables us to construct operator identities between fermionic and bosonic fields.

\subsubsection{Bosonisation identities}

We can expand the boson field compactified to the radius $R$ as
\begin{align}
   \Phi(x)& = {\Phi_0 +\frac{RW}{L}x} -\frac{1}{\sqrt{L}} {\sum_{{n_{k}=-\infty}\atop{n_k\neq0}}^{\infty}}
  \frac{1}{\sqrt{2|k|}}\left(a_{k}+a_{-k}^{\dagger}\right)e^{{\rm i}kx} \label{eq:Phi}
\end{align}
 with $k=\frac{2\pi}{L}n_{k}$ and $\left[a_{k},a_{l}^{\dagger}\right]=\delta_{k,l}$ and $W$ the winding operator. 
It satisfies $\Phi\sim\Phi+2\pi R$; in case of the SG model, a convenient choice for the compactification radius is $R=\sqrt{4\pi}/\beta$). The $k\neq0$ part can be decomposed into two ($\sigma=\pm$) components:
\begin{equation}
    \Phi_\sigma(x)=-\frac{1}{\sqrt{4\pi}}\sum_{n_{k}=1}^{\infty}\frac{1}{\sqrt{n_{k}}}\left(a_{-\sigma k}e^{-\sigma {\rm i}kx}+a_{-\sigma k}^{\dagger}e^{\sigma {\rm i}kx}\right).\label{eq:Phi-chiral}
\end{equation}
The $\sigma=\pm$ components can be easily identified as the left/right moving components of the $\Phi$ field when it is time evolved
under the free massless boson Hamiltonian $H_{CFT}$: \begin{equation}
\mathrm{e}^{+\mathrm{i}H_{CFT}t}\Phi_\sigma(x)\mathrm{e}^{-\mathrm{i}H_{CFT}t}=\Phi_\sigma(x+\sigma t)  \end{equation}
The canonical momentum field $\Pi(x)$ is:
\begin{align}
\Pi(x)= \Pi_0 -\frac{\mathrm{i}}{\sqrt{L}} {\sum_{{n_{k}=-\infty}\atop{n_k\neq0}}^{\infty}}
\sqrt{\frac{|k|}{2}}\left(-a_{k}+a_{-k}^{\dagger}\right)e^{{\rm i}kx}
\end{align}
with $\Pi_0$ the zero mode of the canonical momentum field.

The fermion field
\[
\Psi=\left(\begin{array}{c}
\Psi_{-}\\
\Psi_{+}
\end{array}\right)
\]
satisfying canonical anti-commutation relations and anti-periodic (Neveu-Schwarz) boundary conditions  has the following mode expansion:
\begin{equation}
    \Psi_{\sigma}(x)=
    \frac{1}{\sqrt{L}}
    e^{\sigma{\rm i}\frac{\pi}{L} x}\sum_{n_k=-\infty}^{\infty}\,c_{k,\sigma}e^{-\sigma{\rm i}k x}\label{eq:FermionFieldExpansion}
\end{equation}
with $\sigma=\pm$, $k=\frac{2\pi}{L}n_{k}$ and $\{c_{k,\sigma},c_{l,\rho}^{\dagger}\}=\delta_{\sigma,\rho}\delta_{k,l}$. If evolved with the massless Dirac Hamiltonian $H_{0F} =-\mathrm{i}\int\overline{\Psi}\gamma^{1}\partial_{x}\Psi\,\mathrm{d}x$ the $\sigma=\pm$ modes are the left and right-moving components:
\begin{equation}
\mathrm{e}^{+\mathrm{i}H_{0F}t}\Psi_{\sigma}(x)\mathrm{e}^{-\mathrm{i}H_{0F}t}=\Psi_{\sigma}(x+\sigma t)\,.
\end{equation}
The fermionic number operator is given by:
\begin{equation}
    {N}_{\sigma}\equiv\sum_{n_k=-\infty}^{\infty}\,:\negmedspace c_{k,\sigma}^{\dagger}c_{k,\sigma}\negmedspace:\nonumber
\end{equation}
and acts on the states in $\mathcal{H}_{{N}_{-},{N}_{+}}$ as ${N}_{\sigma}\left|n_{-},n_{+}\right\rangle={n}_{\sigma}\left|n_{-},n_{+}\right\rangle$
Bosonisation is defined by an exact operator identity between boson and fermion fields that follows from \eqref{eq:AprticleHolesBosons}:
\begin{eqnarray}
\Psi_{\sigma}(x)
&=&
    \frac{1}{\sqrt{L}}\,F_{\sigma}\,\mathrm{e}^{-\sigma i\frac{2\pi}{L}\left({N}_{\sigma}-\frac{1}{2}\right)x}\,:\!\mathrm{e}^{-\sigma\mathrm{i}\sqrt{4\pi}\Phi_\sigma(x)}\!:\label{eq:B2Fapp}
\end{eqnarray}
To ensure that the expression in the r.h.s. of the equation acts in the Hilbert space \eqref{eq:HilbertSpaceDecomposition} identically as the fermion field \eqref{eq:FermionFieldExpansion}, it is composed of a bosonic part ($:\!\mathrm{e}^{-\sigma\mathrm{i}\sqrt{4\pi}\Phi_\sigma(x)}\!:$) that acts upon $\mathcal{H}_{\text{Bose}}$ and a part ($F_{\sigma}\,\mathrm{e}^{-i\frac{2\pi}{L}\left({N}_{\sigma}-\frac{1}{2}\right)x}$) that acts upon $\mathcal{H}_{{N}_{-},{N}_{+}}$. The exponentials of the bosonic fields are known as vertex operators, while the operators $F_{\sigma}$ are the Klein factors which act as hopping operators between different $N_\sigma$ sectors:
\[
F_\sigma^{\dagger}\left|n_{-},n_{+}\right\rangle\otimes\left|\psi\right\rangle_{\text{Bose}}\equiv(-1)^{n_{-}+\delta_{\sigma,+}n_{+}}\left|n_{-}+\delta_{\sigma,-},n_{+}+\delta_{\sigma,+}\right\rangle\otimes\left|\psi\right\rangle_{\text{Bose}}
\]
and have the following algebraic properties:
\begin{align}
\left[F_{\sigma},a_{k}^{\dagger}\right] & =\left[F_{\sigma},a_{k}\right]=0\nonumber\\
\left\{ F_{\sigma}^{\dagger},F_{\rho}\right\}  & =2\delta_{\sigma,\rho},\hspace{1em}\forall\sigma,\rho\hspace{1.5cm}(\text{with }F_{\rho}F_{\rho}^{\dagger}=F_{\rho}^{\dagger}F_{\rho}=1)\nonumber\\
\left\{ F_{\sigma}^{\dagger},F_{\rho}^{\dagger}\right\}  & =\left\{ F_{\sigma},F_{\rho}\right\} =0,\hspace{1em}\sigma\neq\rho\nonumber \\
\left[{N}_{\sigma},F_{\rho}^{\dagger}\right] & =\delta_{\sigma,\rho}F_{\rho}^{\dagger},\hspace{1.5cm}\left[{N}_{\sigma},F_{\rho}\right]=-\delta_{\sigma,\rho}F_{\rho}\label{eq:algebra}
\end{align}
where the semicolon denotes the usual normal ordering of the bosonic modes. In terms of the fermions, the bosonic zero modes are given as
\begin{align}
 RW&=\sqrt{\pi}(N_-+N_+), \qquad  \Pi_0 = \frac{\sqrt{\pi}}{L}(-N_-+N_+).
 \\
\end{align}
Formally, the bosonisation identity is proven for systems of finite size $L$. We therefore derive all the expressions in finite volume and take the thermodynamic limit $L\rightarrow\infty$ in the end. 
The inverse relations of (\ref{eq:B2Fapp}) expressing the bosonic fields
in terms of the fermionic ones are given by:
\begin{eqnarray}
\partial_{x}\Phi(x) &=& 
\sqrt{\pi}
\, \sum_{\sigma=\pm} :\negmedspace\Psi_{\sigma}^{\dagger}(x)\Psi_{\sigma}(x)\negmedspace:, \nonumber \\
\Pi(x) &=& 
\sqrt{\pi}
\, \sum_{\sigma=\pm} \sigma :\negmedspace\Psi_{\sigma}^{\dagger}(x)\Psi_{\sigma}(x)\negmedspace:.\label{eq:FermionisationIdentities}
\end{eqnarray}

\subsubsection{Sine-Gordon correlations at the free fermion point} 
Using the above relations we can fermionise the sine-Gordon model \eqref{eq:SGapp} at $\beta=\sqrt{4\pi}$ to obtain the free massive Dirac Hamiltonian \eqref{eq:MFapp}. Exploiting  \eqref{eq:FermionisationIdentities}, the connected correlation functions of the fields ${\partial_x\Phi}$ and ${\Pi}$  can be expressed as:
\begin{align}
\left\langle \Omega\left|\partial_{x}\Phi(x,t)\partial_{y}\Phi(y,t)\right|\Omega\right\rangle_{\text{c}} 
&=
\pi
\sum_{\sigma,\rho=\pm} \bigg(\left\langle \Omega\left|:\negmedspace\Psi_{\sigma}^{\dagger}(x,t)\Psi_{\sigma}(x,t)\negmedspace:\,:\negmedspace\Psi_{\rho}^{\dagger}(y,t)\Psi_{\rho}(y,t)\negmedspace:\right|\Omega\right\rangle-\nonumber\\
&-\left\langle \Omega\left|:\negmedspace\Psi_{\sigma}^{\dagger}(x,t)\Psi_{\sigma}(x,t)\negmedspace:\right|\Omega\right\rangle\left\langle \Omega\left|:\negmedspace\Psi_{\rho}^{\dagger}(y,t)\Psi_{\rho}(y,t)\negmedspace:\right|\Omega\right\rangle\bigg), \label{eq:step1a}
\end{align}
and:
\begin{align}
\left\langle \Omega\left|\Pi(x,t)\Pi(y,t)\right|\Omega\right\rangle_{\text{c}}
&=
\pi\sum_{\sigma,\rho=\pm}\sigma\rho\,\bigg(\left\langle \Omega\left|:\negmedspace\Psi_{\sigma}^{\dagger}(x,t)\Psi_{\sigma}(x,t)\negmedspace:\,:\negmedspace\Psi_{\rho}^{\dagger}(y,t)\Psi_{\rho}(y,t)\negmedspace:\right|\Omega\right\rangle-\nonumber\\
&-\left\langle \Omega\left|:\negmedspace\Psi_{\sigma}^{\dagger}(x,t)\Psi_{\sigma}(x,t)\negmedspace:\right|\Omega\right\rangle\left\langle \Omega\left|:\negmedspace\Psi_{\rho}^{\dagger}(y,t)\Psi_{\rho}(y,t)\negmedspace:\right|\Omega\right\rangle\bigg), \label{eq:step1b}
\end{align}
We also use the identity
\begin{equation}
    \left\langle\Omega\left|:\negmedspace A\negmedspace:\,:\negmedspace B\negmedspace:\right|\Omega\right\rangle-\left\langle \Omega\left|:\negmedspace A\negmedspace:\right|\Omega\right\rangle\left\langle \Omega\left|:\negmedspace B\negmedspace:\right|\Omega\right\rangle=\left\langle \Omega\left|A\,B\right|\Omega\right\rangle-\left\langle \Omega\left|A\right|\Omega\right\rangle\left\langle \Omega\left|B\right|\Omega\right\rangle\label{eq:OmittingNormalOrdering}
\end{equation}
to eliminate the normal ordering of $:\negmedspace\Psi^{\dagger}\Psi\negmedspace:$ pairs in fermionised expressions for correlation functions. This equality follows from $
:\negmedspace A\negmedspace:=A -\left\langle 0\left|A\right|0\right\rangle
$ where $|0\rangle$ is the vacuum state.

\subsection{Computing the dynamics of  correlation functions} 
\subsubsection{Free fermion dynamics} 

As anticipated, using the bosonisation identities (\ref{eq:B2Fapp}) and (\ref{eq:FermionisationIdentities}), the sine-Gordon Hamiltonian at $\beta=\sqrt{4\pi}$ is mapped to the Dirac Hamiltonian \eqref{eq:MFapp}, more precisely in the form:
\[H_{MF}=\sum_{\sigma}\int\mathrm{d}x\left(\sigma\mathrm{i}:\!\Psi_{\sigma}^{\dagger}\partial_{x}\Psi_{\sigma}\!:-M:\!\Psi_{\sigma}^{\dagger}\Psi_{-\sigma}\!:\right)
\]
The equations of motion are:
\[
\dot{\Psi}_{\sigma}=\sigma\partial_{x}\Psi_{\sigma}+\mathrm{i}M\Psi_{-\sigma}
\]
 which can be solved exactly for arbitrary initial conditions. The solution in infinite volume can be expressed in the form where the initial fields are propagated with the retarded Green's functions
\begin{equation}
    \Psi_{\sigma}(x,t)=\sum_{\sigma'=\pm}\int_{-\infty}^{\infty}\mathrm{d}x'\,G_{\sigma\sigma'}(x-x',t)\Psi_{\sigma'}(x')\label{eq:FieldPropagation}
\end{equation}
where 
\begin{align}
G_{\sigma,\sigma'}(x-x',t)&=\Theta(t)\int\frac{\mathrm{d}k}{2\pi}\,\mathrm{e}^{\mathrm{i}k(x-x')}\,\left[\delta_{\sigma,\sigma'}\cos\left(E_{k}t\right)+\left(\sigma\mathrm{i}k\delta_{\sigma,\sigma'}+\mathrm{i}M\delta_{\sigma,-\sigma'}\right)\frac{\sin\left(E_{k}t\right)}{E_{k}}\right] 
\label{eq:Green}
\end{align}
with $E_{k}=\sqrt{k^{2}+M^{2}}$.

Note that the Green's function is Lorentz invariant and it vanishes out of the light-cone, i.e. for $|x-x'|>t$. For $t=0$ it is  $G_{\sigma\sigma'}(x-x',0)=\delta_{\sigma,\sigma'} \delta(x-x')$ and in the special case $M=0$ it reduces to $\delta_{\sigma,\sigma'}\delta(x+\sigma t -x')$ as it should. 

The propagator can be evaluated explicitely \cite{Gradshteyn}:
\begin{align}
 G_{\sigma\sigma'}(x,t) & =
 \frac{1}{2}\Theta(t)\Big(\delta_{\sigma,\sigma'}\left(\partial_t+\sigma\partial_x\right)+ \mathrm{i} M \delta_{\sigma,-\sigma'}\Big)
 \left(\Theta(t^2-x^2) J_0(M\sqrt{t^2-x^2}) \right) \nonumber \\
 & = \Theta(t)\Bigg\{ \frac{1}{2}M\Theta(t^2-x^2) \left[\delta_{\sigma,\sigma'}\left(-t {+}\sigma x \right)\frac{J_1(M\sqrt{t^2-x^2})}{\sqrt{t^2-x^2}} + {\mathrm{i} } \delta_{\sigma,-\sigma'}J_0(M\sqrt{t^2-x^2})\right]  \nonumber \\
 & \qquad \qquad \qquad +\delta_{\sigma,\sigma'}\delta(x+\sigma t) \Bigg\}
\end{align}
where $J_0$ and $J_1$ are the Bessel functions of zeroth and first order respectively.

\subsubsection{Dynamics of correlation functions} 

Substituting \eqref{eq:FieldPropagation} into (\ref{eq:step1a}) and (\ref{eq:step1b}) and using \eqref{eq:OmittingNormalOrdering} yields
\begin{align}
 & \left\langle \Omega\left|\partial_{x}\Phi(x,t)\partial_{y}\Phi(y,t)\right|\Omega\right\rangle_{\text{c}}= \label{eq:step2a}\\
 & 
 \pi
 \sum_{\sigma,\rho,\sigma_{i},\rho_{i}=\pm} \int\mathrm{d}x_{1}\mathrm{d}x_{2}\mathrm{d}y_{1}\mathrm{d}y_{2}\;G_{\sigma\sigma_{1}}^{*}(x-x_{1},t)G_{\sigma\sigma_{2}}(x-x_{2},t)G_{\rho\rho_{1}}^{*}(y-y_{1},t)G_{\rho\rho_{2}}(y-y_{2},t)\times\nonumber\\
 &  \times\;\bigg(\left\langle \Omega\left|\Psi_{\sigma_1}^{\dagger}(x_1)\Psi_{\sigma_2}(x_2)\,\Psi_{\rho_1}^{\dagger}(y_1)\Psi_{\rho_2}(y_2)\right|\Omega\right\rangle-\left\langle \Omega\left|\Psi_{\sigma_1}^{\dagger}(x_1)\Psi_{\sigma_2}(x_2)\right|\Omega\right\rangle\left\langle \Omega\left|\Psi_{\rho_1}^{\dagger}(y_1)\Psi_{\rho_2}(y_2)\right|\Omega\right\rangle\bigg). \nonumber 
\end{align}
and
\begin{align}
 & \left\langle \Omega\left|\Pi(x,t)\Pi(y,t)\right|\Omega\right\rangle_{\text{c}}= \label{eq:step2b}\\
 & \pi \sum_{\sigma,\rho,\sigma_{i},\rho_{i}=\pm}\sigma\rho\int\mathrm{d}x_{1}\mathrm{d}x_{2}\mathrm{d}y_{1}\mathrm{d}y_{2}\;G_{\sigma\sigma_{1}}^{*}(x-x_{1},t)G_{\sigma\sigma_{2}}(x-x_{2},t)G_{\rho\rho_{1}}^{*}(y-y_{1},t)G_{\rho\rho_{2}}(y-y_{2},t)\times\nonumber\\
 &  \times\;\bigg(\left\langle \Omega\left|\Psi_{\sigma_1}^{\dagger}(x_1)\Psi_{\sigma_2}(x_2)\,\Psi_{\rho_1}^{\dagger}(y_1)\Psi_{\rho_2}(y_2)\right|\Omega\right\rangle-\left\langle \Omega\left|\Psi_{\sigma_1}^{\dagger}(x_1)\Psi_{\sigma_2}(x_2)\right|\Omega\right\rangle\left\langle \Omega\left|\Psi_{\rho_1}^{\dagger}(y_1)\Psi_{\rho_2}(y_2)\right|\Omega\right\rangle\bigg). \nonumber 
\end{align}

\begin{figure}[ht]
\centering{\includegraphics[clip,width=0.9\columnwidth]{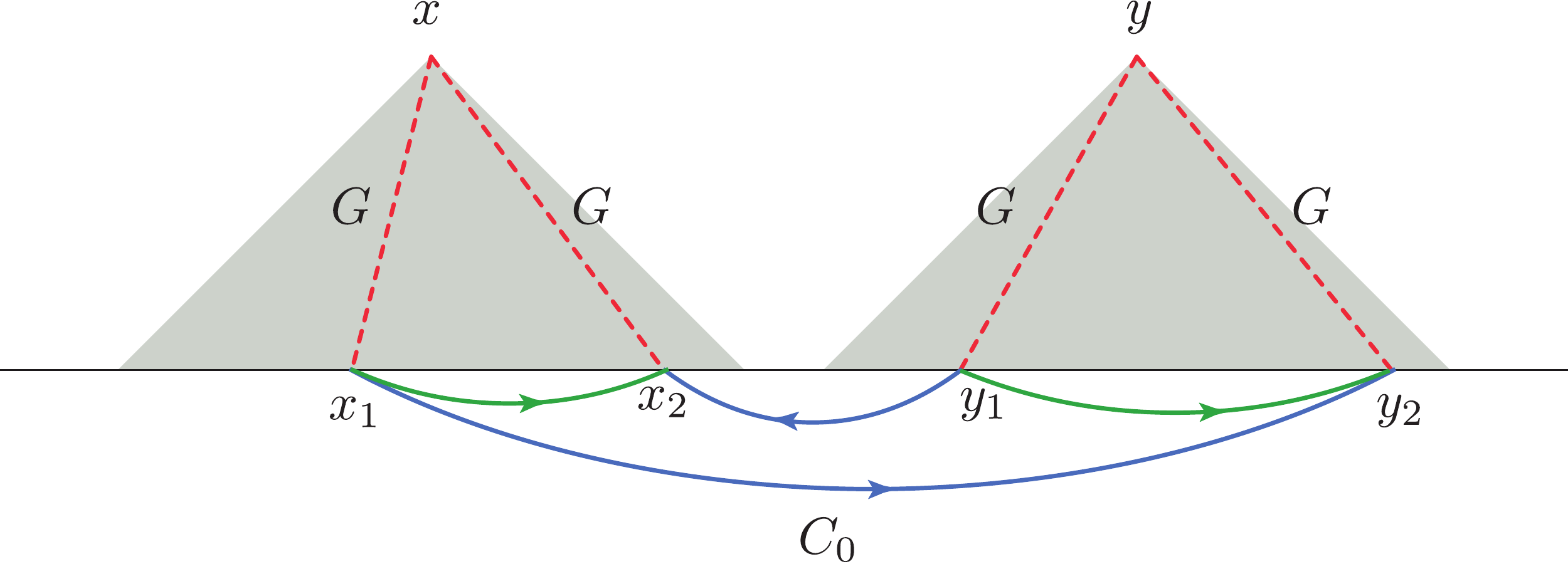}\caption{Diagrammatic representation of the convolution formulas (\ref{eq:step2a}) and (\ref{eq:step2b}) for the derivation of dynamics of correlations at the free fermion point. The dashed red lines denote the propagators, the blue and green lines denote the Klein factor contractions corresponding to the initial fermionic correlations. \label{fig:SM2}}}
\end{figure}

\subsection{Initial fermionic correlation functions}

The last step is to construct the initial fermionic correlation
functions
\begin{eqnarray}
C_{\sigma_{1}\sigma_{2}\rho_{1}\rho_{2}}(x_{1},x_{2},y_{1},y_{2})&\equiv&\left\langle \Omega\left|\Psi_{\sigma_{1}}^{\dagger}(x_{1})\Psi_{\sigma_{2}}(x_{2})\,\Psi_{\rho_{1}}^{\dagger}(y_{1})\Psi_{\rho_{2}}(y_{2})\right|\Omega\right\rangle\nonumber\\
C_{\sigma_{1}\sigma_{2}}(x_{1},x_{2})&\equiv&\left\langle \Omega\left|\Psi_{\sigma_{1}}^{\dagger}(x_{1})\Psi_{\sigma_{2}}(x_{2})\right|\Omega\right\rangle.\label{eq:InitialFermiCorrelations}
\end{eqnarray}
entering (\ref{eq:step2a}) and (\ref{eq:step2b}). 

\subsubsection{Initial state} 
The state $\left|\Omega\right\rangle$ that we are quenching from is the ground state of the Klein-Gordon model
\begin{equation}
H_{KG}=\int\left(\frac{1}{2}\Pi{}^{2}+\frac{1}{2}(\partial_{x}\Phi)^{2}+\frac{1}{2}m_{0}^{2}\Phi^{2}\right)\mathrm{d}x\nonumber.
\end{equation}
When embedded in the fermionic Hilbert space \eqref{eq:HilbertSpaceDecomposition} the state $\left|\Omega\right\rangle$ only has excitations of bosonic particle-hole type:
\[
\left|\Omega\right\rangle_{\mathcal{H}_{\text{Fermi}}}=\left|0,0\right\rangle\otimes\left|\Omega\right\rangle.
\]
In \eqref{eq:step1a},\eqref{eq:step1b} and all other fermionic expressions we drop the index $\mathcal{H}_{\text{Fermi}}$ for simplicity of notation but always have $\left|\Omega\right\rangle_{\mathcal{H}_{\text{Fermi}}}$ in mind when we write $\left|\Omega\right\rangle$. 

\subsubsection{Fermionic superselection rules}  
Using the bosonisation identity (\ref{eq:B2Fapp}), the initial fermionic correlations \eqref{eq:InitialFermiCorrelations} can be expressed in terms of bosonic fields.  Collecting together all Klein factors entering in $C_{\sigma_{1}\sigma_{2}\rho_{1}\rho_{2}}(x_{1},x_{2},y_{1},y_{2})$ and $C_{\sigma_{1}\sigma_{2}}(x_{1},x_{2})$ using the algebraic relations (\ref{eq:algebra}) and commuting the Klein factors past the $\mathrm{e}^{-i\frac{2\pi}{L}\left({N}_{\sigma}-\frac{1}{2}\right)x}$ factors from $\eqref{eq:B2Fapp}$ using the identity $\mathrm{e}^{A}B=B\mathrm{e}^{A+c}$ if $\left[A,B\right]=cB$ for $c\in\mathbb{C}$ gives the following phases: 
\begin{align}
&\Theta_{\sigma_{1}\sigma_{2}\rho_{1}\rho_{2}}(x_{1},x_{2},y_{1},y_{2})\nonumber\\
&\equiv\mathrm{e}^{i\frac{2\pi}{L}\left[\sigma_{1}\left(\frac{1}{2}-\delta_{\sigma_{1},\sigma_{2}}+\delta_{\sigma_{1},\rho_{1}}-\delta_{\sigma_{1},\rho_{2}}\right)x_{1}-\sigma_{2}\left(-\frac{1}{2}+\delta_{\sigma_{2},\rho_{1}}-\delta_{\sigma_{2},\rho_{2}}\right)x_{2}+\rho_{1}\left(\frac{1}{2}-\delta_{\rho_{1},\rho_{2}}\right)y_{1}+\frac{1}{2}\rho_{2}y_{2}\right]} 
\end{align}
for $C_{\sigma_{1}\sigma_{2}\rho_{1}\rho_{2}}$ and 
and
\begin{align}
\Theta_{\sigma_{1}\sigma_{2}}(x_{1},x_{2})\equiv\mathrm{e}^{i\frac{2\pi}{L}\left[\sigma_{1}\left(\frac{1}{2}-\delta_{\sigma_{1},\sigma_{2}}\right)x_{1}+\frac{1}{2}\sigma_{2}x_{2}\right]}
\end{align}
for $C_{\sigma_{1}\sigma_{2}}$. Both of these phases become identically equal to 1 once we take the thermodynamic limit $L\rightarrow\infty$.

The Klein factors impose the  following superselection rules when contracted with the $\mathcal{H}_{\hat{{N}}}$ part of $\left|\Omega\right\rangle$:
\begin{align}
\left\langle0,0\right| F_{\sigma_{1}}^{\dagger}F_{\sigma_{2}}F_{\rho_{1}}^{\dagger}F_{\rho_{2}}\left|0,0\right\rangle=\delta_{\sigma_{1},\sigma_{2}}\delta_{\rho_{1},\rho_{2}}+\delta_{\sigma_{1},\rho_{2}}\delta_{\sigma_{2},\rho_{1}}(1-\delta_{\sigma_{1}\sigma_{2}}).
\end{align}
This means that any fermion $F_{\sigma}$ appearing in the expression must be matched with an antifermion $F_{\sigma}^{\dagger}$ of the same type $\sigma$, or in other words the string of operators must preserve the total $N_{+}=0$ and $N_{-}=0$. Therefore the only non-vanishing combinations are $C_{++++},C_{++--},C_{+--+}$ and those with all signs reversed. For the two-point functions we have:
\begin{align}
\left\langle0,0\right| F_{\sigma_1}^{\dagger}F_{\sigma_2}\left|0,0\right\rangle=\delta_{\sigma_1,\sigma_2}
\end{align}
so the only nonvanishing combinations are $C_{++}$ and $C_{--}$.

\subsubsection{Contracting vertex operators} 
We therefore end up with the following expressions:
\begin{align}
&C_{\sigma_{1}\sigma_{2}\rho_{1}\rho_{2}}(x_{1},x_{2},y_{1},y_{2})=
\frac{1}{L^2}
\bigg\{\delta_{\sigma_{1},\sigma_{2}}\delta_{\rho_{1},\rho_{2}}+\delta_{\sigma_{1},\rho_{2}}\delta_{\sigma_{2},\rho_{1}}(1-\delta_{\sigma_{1}\sigma_{2}})\bigg\}\,\Theta_{\sigma_{1}\sigma_{2}\rho_{1}\rho_{2}}(x_{1},x_{2},y_{1},y_{2})\times\nonumber\\
&\times\left\langle \Omega\left|:\negmedspace\mathrm{e}^{+\sigma_{1}\sqrt{4\pi}\mathrm{i}\Phi_{\sigma_{1}}(x_{1})}\negmedspace:\,:\negmedspace\mathrm{e}^{-\sigma_{2}\sqrt{4\pi}\mathrm{i}\Phi_{\sigma_{2}}(x_{2})}\negmedspace:\,:\negmedspace\mathrm{e}^{+\rho_{1}\sqrt{4\pi}\mathrm{i}\Phi_{\rho_{1}}(y_{1})}\negmedspace:\,:\negmedspace\mathrm{e}^{-\rho_{2}\sqrt{4\pi}\mathrm{i}\Phi_{\rho_{2}}(y_{2})}\negmedspace:\right|\Omega\right\rangle \nonumber\\
&C_{\sigma_{1}\sigma_{2}}(x_{1},x_{2})=
\frac{1}{L}
\,\delta_{\sigma_1,\sigma_2}\,\Theta_{\sigma_{1}\sigma_{2}}(x_{1},x_{2})\,\left\langle \Omega\left|:\negmedspace\mathrm{e}^{+\sigma_{1}\sqrt{4\pi}\mathrm{i}\Phi_{\sigma_{1}}(x_{1})}\negmedspace:\,:\negmedspace\mathrm{e}^{-\sigma_{2}\sqrt{4\pi}\mathrm{i}\Phi_{\sigma_{2}}(x_{2})}\negmedspace:\right|\Omega\right\rangle\label{eq:step3a}
\end{align}
In order to evaluate these correlation functions we exploit the fact that, since the initial state $|\Omega\rangle$ is the ground state of Klein Gordon model,
it is Gaussian in terms of the bosonic field. For such Gaussian states
Wick's theorem yields
\begin{align}
 & \left\langle \prod_{i}\mathrm{e}^{\mathrm{i}a_{i}\Phi_{\sigma_{i}}(x_{i})}\right\rangle =\mathrm{e}^{-\frac{1}{2}\sum_{i}a_{i}^{2}\left\langle \Phi_{\sigma_{i}}^{2}(x_{i})\right\rangle -\sum_{i<j}a_{i}a_{j}\left\langle \Phi_{\sigma_{i}}(x_{i})\Phi_{\sigma_{j}}(x_{j})\right\rangle }\label{eq:gaussian}
\end{align}
Using the Baker-Campbell-Hausdorff formula we obtain an analogous formula for normal-ordered exponentials
\begin{align}
 & \left\langle \prod_{i}:\!\mathrm{e}^{\mathrm{i}a_{i}\Phi_{\sigma_{i}}(x_{i})}\!:\right\rangle =\mathrm{e}^{-\frac{1}{2}\sum_{i}a_{i}^{2}\left\langle :\Phi_{\sigma_{i}}^{2}(x_{i}):\right\rangle -\sum_{i<j}a_{i}a_{j}\left\langle \Phi_{\sigma_{i}}(x_{i})\Phi_{\sigma_{j}}(x_{j})\right\rangle }\label{eq:gaussian2}
\end{align}
Applying it to our problem results in
\begin{align}
 & \left\langle \Omega\left|:\negmedspace\mathrm{e}^{+\sigma_{1}\sqrt{4\pi}\mathrm{i}\Phi_{\sigma_{1}}(x_{1})}\negmedspace:\,:\negmedspace\mathrm{e}^{-\sigma_{2}\sqrt{4\pi}\mathrm{i}\Phi_{\sigma_{2}}(x_{2})}\negmedspace:\,:\negmedspace\mathrm{e}^{+\sigma_{3}\sqrt{4\pi}\mathrm{i}\Phi_{\rho_{1}}(x_3)}\negmedspace:\,:\negmedspace\mathrm{e}^{-\rho_{2}\sqrt{4\pi}\mathrm{i}\Phi_{\sigma_4}(x_4)}\negmedspace:\right|\Omega\right\rangle = \nonumber \\
 & \qquad=\exp\Bigg[-2\pi\sum_{i=1}^{4}\left\langle :\!\Phi_{\sigma_{i}}^{2}\!:\right\rangle_{\Omega} - \nonumber \\
 & \qquad - 4\pi\Big(-\sigma_{1}\sigma_{2}\left\langle \Phi_{\sigma_{1}}(x_{1})\Phi_{\sigma_{2}}(x_{2})\right\rangle_{\Omega} -\sigma_{1}\sigma_{4}\left\langle \Phi_{\sigma_{1}}(x_{1})\Phi_{\sigma_{4}}(x_{4})\right\rangle_{\Omega} -\sigma_{2}\sigma_{3}\left\langle \Phi_{\sigma_{2}}(x_{2})\Phi_{\sigma_{3}}(x_{3})\right\rangle_{\Omega} \nonumber \\
 & \qquad\qquad-\sigma_{3}\sigma_{4}\left\langle \Phi_{\sigma_{3}}(x_{3})\Phi_{\sigma_{4}}(x_{4})\right\rangle_{\Omega} +\sigma_{1}\sigma_{3}\left\langle \Phi_{\sigma_{1}}(x_{1})\Phi_{\sigma_{3}}(x_{3})\right\rangle_{\Omega} +\sigma_{2}\sigma_{4}\left\langle \Phi_{\sigma_{2}}(x_{2})\Phi_{\sigma_{4}}(x_{4})\right\rangle_{\Omega} \Big)\Bigg]
 \nonumber \\
&  \left\langle \Omega\left|:\negmedspace\mathrm{e}^{+\sigma_{1}\sqrt{4\pi}\mathrm{i}\Phi_{\sigma_{1}}(x_{1})}\negmedspace:\,:\negmedspace\mathrm{e}^{-\sigma_{2}\sqrt{4\pi}\mathrm{i}\Phi_{\sigma_{2}}(x_{2})}\negmedspace:\right|\Omega\right\rangle \nonumber \\
& \qquad = \exp\Bigg[-2\pi\sum_{i=1}^{2}\left\langle :\!\Phi_{\sigma_{i}}^{2}\!:\right\rangle_{\Omega} +4\pi\sigma_{1}\sigma_{2}\left\langle \Phi_{\sigma_{1}}(x_{1})\Phi_{\sigma_{2}}(x_{2})\right\rangle_{\Omega}\Bigg]\label{eq:step3b}
\end{align}
where we abbreviated $\left\langle \Omega\left|\,\bullet\, \right|\Omega\right\rangle$ by  $\left\langle \,\bullet\, \right\rangle_{\Omega}$.

\subsubsection{Initial two-point functions of bosonic fields}

The problem now reduces to calculating the initial two-point correlation functions $\left\langle \Omega\left| \Phi_{\sigma_1}(x_1)\Phi_{\sigma_2}(x_2)\right|\Omega\right\rangle$,
which are correlations between the chiral components of the bosonic
field evaluated in the Klein-Gordon ground state. From 
(\ref{eq:Phi-chiral}) we find
\begin{align}
 & \left\langle \Omega\left| \Phi_{\sigma_1}(x_1)\Phi_{\sigma_2}(x_2)\right|\Omega\right\rangle \nonumber \\
 & =\frac{1}{L}\sum_{n_k=1}^{\infty}\Bigg\{\delta_{\sigma_1,\sigma_2}\left[\frac{1}{2k}\mathrm{e}^{-\sigma_1\mathrm{i}k(x_1-x_2)}+\frac{1}{4k}\left(\frac{E_{0k}}{k}+\frac{k}{E_{0k}}-2\right)\cos k(x_1-x_2)\right]\nonumber \\
 & \quad-\delta_{\sigma_1,-\sigma_2}\frac{1}{4k}\left(\frac{E_{0k}}{k}-\frac{k}{E_{0k}}\right)\cos k(x_1-x_2)\Bigg\}\label{eq:boson_cf}
\end{align}
where $E_{0k}=\sqrt{k^{2}+m_{0}^{2}}$ and the sum runs over discrete momenta, $k=2\pi n_k/L$,
for positive integers $n_k$. It is easy to verify that for $m_{0}=0$
the only term that does not vanish is the first one: this equals $-\frac{1}{4\pi}\log\left(1-\mathrm{e}^{-\sigma\mathrm{i}2\pi x/L}\right)$
and is the one that results in the standard CFT ground state correlations.

\subsubsection{Putting the building blocks together} 

Substituting \eqref{eq:boson_cf} in (\ref{eq:step3b}) and then back to (\ref{eq:step3a})
we finally find explicit formulae for the non-vanishing initial four-point fermionic correlations
\begin{align}
&C_{\sigma\sigma\sigma\sigma}(x_{1},x_{2},y_{1},y_{2}) =  \Theta_{\sigma\sigma\sigma\sigma}(x_{1},x_{2},y_{1},y_{2})\,\,C_{\sigma\sigma\sigma\sigma}^{0}(x_{1},x_{2},y_{1},y_{2})\times\nonumber\\
&\times\exp\Big[4\pi\Big(I_{1}(x_{1}-x_{2})+I_{1}(y_{1}-y_{2})+I_{1}(x_{1}-y_{2})+I_{1}(x_{2}-y_{1})-I_{1}(x_{1}-y_{1})-I_{1}(x_{2}-y_{2})\Big)\Big]\nonumber \\
&C_{\sigma\sigma(-\sigma)(-\sigma)}(x_{1},x_{2},y_{1},y_{2}) = \Theta_{\sigma\sigma(-\sigma)(-\sigma)}(x_{1},x_{2},y_{1},y_{2})\,\,C_{\sigma\sigma(-\sigma)(-\sigma)}^{0}(x_{1},x_{2},y_{1},y_{2})\times\nonumber\\
&\times\exp\Big[4\pi\Big(I_{1}(x_{1}-x_{2})+I_{1}(y_{1}-y_{2})+I_{2}(x_{1}-y_{2})+I_{2}(x_{2}-y_{1})-I_{2}(x_{1}-y_{1})-I_{2}(x_{2}-y_{2})\Big)\Big]\nonumber \\
&C_{\sigma(-\sigma)(-\sigma)\sigma}(x_{1},x_{2},y_{1},y_{2}) = \Theta_{\sigma(-\sigma)(-\sigma)\sigma}(x_{1},x_{2},y_{1},y_{2})\,\,C_{\sigma(-\sigma)(-\sigma)\sigma}^{0}(x_{1},x_{2},y_{1},y_{2})\times\nonumber\\
&\times\exp\Big[4\pi\Big(I_{2}(x_{1}-x_{2})+I_{2}(y_{1}-y_{2})+I_{1}(x_{1}-y_{2})+I_{1}(x_{2}-y_{1})-I_{2}(x_{1}-y_{1})-I_{2}(x_{2}-y_{2})\Big)\Big]\label{eq:fermion_cf}
\end{align}
where $C_{\sigma_{1}\sigma_{2}\rho_{1}\rho_{2}}^{0}(x_{1},x_{2},y_{1},y_{2})$
denotes the CFT part of the corresponding correlations:
\begin{align}
&C_{\sigma\sigma\sigma\sigma}^{0}(x_{1},x_{2},y_{1},y_{2})
= \frac{1}{L^2}
\exp\Big[4\pi\Big(I_{0}^{\sigma}(x_{1}-x_{2})+I_{0}^{\sigma}(y_{1}-y_{2})+I_{0}^{\sigma}(x_{1}-y_{2})\nonumber \\
&\qquad +I_{0}^{\sigma}(x_{2}-y_{1})-I_{0}^{\sigma}(x_{1}-y_{1})-I_{0}^{\sigma}(x_{2}-y_{2})\Big)\Big]\nonumber \\
&C_{\sigma\sigma(-\sigma)(-\sigma)}^{0}(x_{1},x_{2},y_{1},y_{2}) =
\frac{1}{L^2}
\exp\Big[4\pi\Big(I_{0}^{\sigma}(x_{1}-x_{2})+I_{0}^{-\sigma}(y_{1}-y_{2})\Big)\Big]\nonumber \\
&C_{\sigma(-\sigma)(-\sigma)\sigma}^{0}(x_{1},x_{2},y_{1},y_{2}) =
\frac{1}{L^2}
\exp\Big[4\pi\Big(I_{0}^{\sigma}(x_{1}-y_{2})+I_{0}^{-\sigma}(x_{2}-y_{1})\Big)\Big]\label{eq:fermion_cf_0}.
\end{align}
The non-vanishing two-point correlations are
\begin{align}
C_{\sigma\sigma}(x_{1},x_{2}) & =  \Theta_{\sigma\sigma}(x_{1},x_{2})\, C_{\sigma\sigma}^{0}(x_{1},x_{2})\,\exp\Big[4\pi I_{1}(x_{1}-x_{2})\Big],\label{eq:fermion_cf_2pt}
\end{align}
with 
\[
C_{\sigma\sigma}^{0}(x_{1},x_{2})= 
\frac{1}{L} 
\exp\Big[4\pi I_{0}^{\sigma}(x_{1}-x_{2})\Big].
\]

The functions $I_{0}^{\pm},I_{1}$ and $I_{2}$ are given by
\begin{align}
I_{1}(x) & \coloneqq\frac{1}{L}\sum_{n_k=1}^{\infty}\frac{1}{4k}\left(\frac{E_{0k}}{k}+\frac{k}{E_{0k}}-2\right)\left(\cos kx-1\right)\nonumber\\
I_{2}(x) & \coloneqq\frac{1}{L}\sum_{n_k=1}^{\infty}\frac{1}{4k}\left(\frac{E_{0k}}{k}-\frac{k}{E_{0k}}\right)\left(\cos kx-1\right)\nonumber\\
I_{0}^{\sigma}(x) & \coloneqq\frac{1}{L}\sum_{n_k=1}^{\infty}\frac{1}{2k}\mathrm{e}^{-\sigma\mathrm{i}kx}\label{eq:integrals}
\end{align}
It can be verified that the above formulae reproduce the known fermionic correlations for the massless free fermion (CFT) ground state. In particular, the fermionic antisymmety property is granted by the canonical anticommutation relations of the fermion field as defined by \eqref{eq:B2Fapp}. The $I_{0}^{\sigma}(x)$, when evaluated, gives terms $\propto\log L$ that cancel the $\frac{1}{L}$ factors in front of the initial correlations.

The thermodynamic limit $L\rightarrow\infty$ is obtained by  replacing the sums in the functions \eqref{eq:integrals} by integrals:
\begin{equation}
\frac{1}{L}\sum_{n_k=1}^{\infty}\rightarrow\,\int_{0}^{\infty}\frac{\mathrm{d}k}{2\pi}\,
\end{equation}
except when there is an infrared singularity, while the phases $\Theta_{\sigma_{1}\sigma_{2}\rho_{1}\rho_{2}}$ and $\Theta_{\sigma_{1}\sigma_{2}}$ become identically equal to one. 
 
The integrals corresponding to  $I_{1}(x)$ and
$I_{2}(x)$ are infrared and ultraviolet convergent and can easily be evaluated numerically. The expression $I_{0}^\sigma(x)$ has an infrared divergence and should be kept as a discrete until the end of the calculation. This part is the only one that does not vanish in the case $m_0=m$ and it is responsible for the CFT ground state correlations. Specifically it can be shown that upon exponentiation 
\begin{equation}
\lim_{L\to\infty}\frac{1}{L}\exp{(4\pi I_{0}^\sigma(x))}=-\frac{\sigma\mathrm{i}}{2\pi}\frac{1}{x},
\end{equation}
which gives the standard algebraically decaying CFT correlations
\begin{equation}
C_{\sigma\sigma}^{0}(x_{1},x_{2})= -\frac{\sigma\mathrm{i}}{2\pi}\frac{1}{x_1-x_2}
\end{equation}
It is worth noticing that, while  both
$I_{1}(x)$ and $I_{2}(x)$ vanish for $|x|\to0$ and therefore do not alter the
short distance behaviour of any correlator (which should indeed be
controlled by the CFT scaling laws), at large distances their
asymptotic behaviour is
\begin{align}
I_{1}(x) & \sim-\frac{m_{0}|x|}{16}+\frac{1}{4\pi}\log\left|x\right|+c_1\nonumber\\
I_{2}(x) & \sim-\frac{m_{0}|x|}{16}+c_2\label{eq:integrals-asympt}
\end{align}
where $c_1,c_2$ are numerical constants. These asymptotics are precisely cancel the algebraic decay of CFT correlations and switch it to an exponential one, as expected for a massive KG ground state $\left|\Omega\right\rangle$. 

The initial fermionic correlations can be computed by combining (\ref{eq:fermion_cf}), (\ref{eq:fermion_cf_2pt}) and (\ref{eq:integrals}). Substituting
in (\ref{eq:step2a}) and (\ref{eq:step2b}) and using (\ref{eq:Green}) gives
the time evolution of the two-point connected correlation function
$C_{\partial\Phi}(x,y;t)$. The result of a numerical integration is
shown in Fig.~\ref{fig:2a} of the main text. 
Note that while the numerical computation is efficient outside of the horizon (for $|x-y|>2t$), this is not the case inside the horizon (for $|x-y|<2t$) due to the presence of singularities at $x_1=y_2$ and $x_2=y_1$. In this region it is necessary to split the integrands into their singular and non-singular parts and evaluate them separately, noticing that the singular part can be expressed as a sum of products of double integrals (instead of quadruple), which speeds up its evaluation. Moreover a short-distance cutoff  $\epsilon\approx0.1/m_0$ has been used to smear the singularities and make the numerical integration feasible.

\subsection{Explanation of the out-of-horizon effect}\label{app:asympt}
The asymptotics of connected correlations
at any time $t$ and large distance $r=|x-y|$ can be derived using  (\ref{eq:fermion_cf}), (\ref{eq:fermion_cf_2pt}),
(\ref{eq:integrals}), (\ref{eq:step2a}), (\ref{eq:step2b}) and (\ref{eq:Green}). From (\ref{eq:step2a}), (\ref{eq:step2b}) and the fact that the fermionic propagators have support only inside the light-cones it follows that the large distance asymptotics outside of 
the horizon lines $r=2t$ are determined by the asymptotics of the
initial four-point fermionic correlations as the distance between
the pairs of coordinates $x_{1},x_{2}$ and $y_{1},y_{2}$ becomes
large. The asymptotic expansions  (\ref{eq:integrals-asympt}) imply that the cross-terms $C_{\sigma(-\sigma)(-\sigma)\sigma}$
do not tend to $C_{\sigma(-\sigma)}C_{(-\sigma)\sigma}$ (which vanishes due to the Klein
factor superselection rules), but they have a non-zero limit instead
\begin{align}
\lim_{r\to\pm\infty} & C_{\sigma(-\sigma)(-\sigma)\sigma}(x_{1},x_{2},y_{1}+r,y_{2}+r) = \nonumber \\ &
\tfrac{A}{(2\pi)^{2}} \, \mathrm{e}^{4\pi I_{2}(x_{1}-x_{2})+4\pi I_{2}(y_{1}-y_{2})}
\neq C_{\sigma(-\sigma)}(x_{1},x_{2})C_{(-\sigma)\sigma}(y_{1},y_{2})=0\label{eq:initial_asympt}
\end{align}
where $A=\exp[8\pi(c_1-c_2)]$  and $c_1,c_2$ are the constants defined in (\ref{eq:integrals-asympt}). 
This is in contrast with the behaviour of $C_{\sigma\sigma\sigma\sigma}$ and $C_{\sigma\sigma(-\sigma)(-\sigma)}$
correlations, which factorise at large distances to $C_{\sigma\sigma}C_{\sigma\sigma}$
and $C_{\sigma\sigma}C_{(-\sigma)(-\sigma)}$ respectively.

The above violation of clustering implies the dynamical emergence
of infinite range correlations presented in the main text. More specifically,
from (\ref{eq:step2a}), (\ref{eq:step2b}) and (\ref{eq:initial_asympt}) the asymptotic
values of the connected correlation functions $C_{\partial\Phi}(x,y;t)$ and $C_{\Pi}(x,y;t)$ at large distance $r=|x-y|$ are
\begin{align}
\lim_{r\to\infty}C_{\partial\Phi}(0,r;t) & = 0 \\
\lim_{r\to\infty}C_{\Pi}(0,r;t) & =\tfrac{A}{2\pi} \left|\sum_{\sigma}\sigma\int dx_{1}dx_{2}\,G_{\sigma,+1}^{*}(x_{1},t)G_{\sigma,-1}(x_{2},t)\,\mathrm{e}^{4\pi I_{2}(x_{1}-x_{2})}\right|^{2}\label{eq:result}
\end{align}
which is plotted in Fig.~\ref{fig:2b} of the main text. At $t=0$ the Green's functions \eqref{eq:Green} corresponding to the time evolution of the cross-terms  vanish, so these terms do not contribute to the initial state. For $t>0$ they are propagated by the off-diagonal part of the Green's functions. As we can easily see by means of the stationary phase method, the above expression is oscillatory and decays with time as 
\begin{align}
\lim_{r\to\infty}C_{\Pi}(0,r;t) \sim \frac{AB^2M}{(4\pi)^2} \frac{(1+\sin4Mt)}{t}
\quad \text{for }t\to\infty
\label{eq:asympt}
\end{align}
with $B=\int_{-\infty}^{+\infty}\mathrm{e}^{4\pi I_{2}(s)} \mathrm{d}s$. We remark that the leading correction to this asymptotic value at large distances turns out to decay exponentially with the distance.

\end{document}